\documentclass{aastex61}
\usepackage[utf8]{inputenc}
\usepackage{amsmath,amssymb}
\usepackage{longtable}
\usepackage{graphicx}
\usepackage{listings}
\usepackage{hyperref}

\newcommand{\teff}{\ensuremath{T_{\rm eff}}}
\newcommand{\logg}{\ensuremath{\log g}}
\newcommand{\tmag}{\ensuremath{T}}

\newcommand{\msun}{\ensuremath{M_\odot}}
\newcommand{\rsun}{\ensuremath{R_\odot}}
\newcommand{\lbol}{\ensuremath{L_{\rm bol}}}
\newcommand{\rearth}{\ensuremath{R_{\oplus}}}

\newcommand{\gbp}{\ensuremath{G_{\rm BP}}}
\newcommand{\grp}{\ensuremath{G_{\rm RP}}}

\definecolor{mygreen}{rgb}{0.0, 0.5, 0.0}
\definecolor{mylila}{rgb}{0.5, 0.0, 0.5}
\definecolor{myblue}{rgb}{0.0, 0.0, 0.5}
\definecolor{mypurple}{rgb}{0.58, 0.42, 0.87}

\begin{document}

\title{The Revised {\it TESS\/} Input Catalog and Candidate Target List}
\author[0000-0002-3481-9052]{Keivan G.\ Stassun}
\affiliation{Vanderbilt University, Department of Physics \& Astronomy, 6301 Stevenson Center Ln., Nashville, TN 37235, USA}
\affiliation{Fisk University, Department of Physics, 1000 18th Ave. N., Nashville, TN 37208, USA}

\author[0000-0002-0582-1751]{Ryan J.\ Oelkers}
\affiliation{Vanderbilt University, Department of Physics \& Astronomy, 6301 Stevenson Center Ln., Nashville, TN 37235, USA}

\author[0000-0001-8120-7457]{Martin Paegert}
\affiliation{Center for Astrophysics $\vert$ Harvard \& Smithsonian, 60 Garden Street, Cambridge, MA 02138, USA}

\author[0000-0002-5286-0251]{Guillermo Torres}
\affiliation{Center for Astrophysics $\vert$ Harvard \& Smithsonian, 60 Garden Street, Cambridge, MA 02138, USA}

\author{Joshua Pepper}
\affiliation{Lehigh University, Department of Physics, 16 Memorial Drive East, Bethlehem, PA, 18015, USA}

\author[0000-0002-3657-0705]{Nathan De Lee} 
\affiliation{1 Department of Physics, Geology, and Engineering Technology, Northern Kentucky University, Highland Heights, KY 41099, USA}

\author{Kevin Collins}
\affiliation{Vanderbilt University, Department of Physics \& Astronomy, 6301 Stevenson Center Ln., Nashville, TN 37235, USA}

\author[0000-0001-9911-7388]{David W.\ Latham}
\affiliation{Center for Astrophysics $\vert$ Harvard \& Smithsonian, 60 Garden Street, Cambridge, MA 02138, USA}





\author[0000-0002-0638-8822]{Philip S.\ Muirhead}
\affiliation{Department of Astronomy, Institute for Astrophysical Research, Boston University, 725 Commonwealth Ave., Boston, MA 02215, USA}

\author{Jay Chittidi}
\affiliation{Department of Physics \& Astronomy, Vassar College, Poughkeepsie, NY 12604, USA}
\affiliation{Maria Mitchell Observatory, 4 Vestal Street, Nantucket, MA 02554, USA}

\author[0000-0002-0149-1302]{B\'arbara Rojas-Ayala}
\affiliation{Departamento de Ciencias Fisicas, Universidad Andres Bello, Fernandez Concha 700, Las Condes, Santiago, Chile}


\author[0000-0003-0556-027X]{Scott W.\ Fleming}
\affiliation{Space Telescope Science Institute, 3700 San Martin Drive, Baltimore, MD 21218, USA}

\author{Mark E.\ Rose}
\affiliation{Leidos, Inc.}
\affiliation{NASA Ames Research Center, Moffett Field, CA 94035, USA}

\author{Peter Tenenbaum}
\affiliation{SETI Institute}
\affiliation{NASA Ames Research Center, Moffett Field, CA 94035, USA}

\author{Eric B.\ Ting}
\affiliation{NASA Ames Research Center, Moffett Field, CA 94035, USA}

\author[0000-0002-7084-0529]{Stephen R.\ Kane}
\affiliation{Department of Earth and Planetary Sciences, University of California, Riverside, CA 92521, USA}


\author[0000-0001-7139-2724]{Thomas Barclay}
\affiliation{NASA Goddard Space Flight Center, 8800 Greenbelt Rd, Greenbelt, MD 20771}
\affiliation{University of Maryland, Baltimore County, 1000 Hilltop Cir, Baltimore, MD 21250}

\author{Jacob L.\ Bean}
\affiliation{Department of Astronomy \& Astrophysics, University of
Chicago, Chicago, IL 60637, USA}

\author{C.~E.\ Brassuer}
\affiliation{Space Telescope Science Institute, 3700 San Martin Drive, Baltimore, MD 21218, USA}

\author{David Charbonneau}
\affiliation{Center for Astrophysics $\vert$ Harvard \& Smithsonian, 60 Garden Street, Cambridge, MA 02138, USA}

\author{Jack J.\ Lissauer}
\affiliation{NASA Ames Research Center, Moffett Field, CA 94035, USA}

\author[0000-0003-3654-1602]{Andrew W.\ Mann}
\affiliation{Department of Physics and Astronomy, University of North Carolina at Chapel Hill, Chapel Hill, NC 27599, USA}

\author[0000-0002-8058-643X]{Brian McLean}
\affiliation{Space Telescope Science Institute, 3700 San Martin Drive, Baltimore, MD 21218, USA}

\author[0000-0001-7106-4683]{Susan Mulally}
\affiliation{Space Telescope Science Institute, 3700 San Martin Drive, Baltimore, MD 21218, USA}

\author[0000-0001-8511-2981]{Norio Narita}
\affiliation{Astrobiology Center, 2-21-1 Osawa, Mitaka, Tokyo 181-8588, Japan}
\affiliation{JST, PRESTO, 2-21-1 Osawa, Mitaka, Tokyo 181-8588, Japan}
\affiliation{National Astronomical Observatory of Japan, 2-21-1 Osawa, Mitaka, Tokyo 181-8588, Japan}
\affiliation{Instituto de Astrof\'{i}sica de Canarias (IAC), 38205 La Laguna, Tenerife, Spain}

\author[0000-0002-8864-1667]{Peter Plavchan}
\affiliation{Department of Physics \& Astronomy, MS 3F3; George Mason University; 4400 University Drive; Fairfax, VA 22030, USA}

\author{George R.\ Ricker}
\affiliation{Kavli Institute for Astrophysics and Space Research, Massachusetts Institute of Technology, Cambridge, MA 02139, USA}

\author[0000-0001-7014-1771]{Dimitar Sasselov}
\affiliation{Center for Astrophysics $\vert$ Harvard \& Smithsonian, 60 Garden Street, Cambridge, MA 02138, USA}

\author[0000-0002-6892-6948]{S.~Seager}
\affiliation{Kavli Institute for Astrophysics and Space Research, Massachusetts Institute of Technology, Cambridge, MA 02139, USA}
\affiliation{Department of Physics, Massachusetts Institute of Technology, Cambridge, MA 02139, USA}
\affiliation{Department of Earth, Atmospheric and Planetary Sciences, Massachusetts Institute of Technology, Cambridge, MA 02139, USA}
\affiliation{Department of Aeronautics and Astronautics, MIT, 77 Massachusetts Avenue, Cambridge, MA 02139, USA}

\author[0000-0002-0920-809X]{Sanjib Sharma}
\affiliation{Sydney Institute for Astronomy(SIfA), School of Physics, University of Sydney, NSW 2006, Australia}

\author{Bernie Shiao}
\affiliation{Space Telescope Science Institute, 3700 San Martin Drive, Baltimore, MD 21218, USA}

\author[0000-0002-7504-365X]{Alessandro Sozzetti}
\affiliation{INAF - Osservatorio Astrofisico di Torino, Via Osservatorio 20, 10025 PIno Torinese, Italy}

\author[0000-0002-4879-3519]{Dennis Stello}
\affiliation{School of Physics, University of New South Wales, NSW 2052, Australia}
\affiliation{Sydney Institute for Astronomy(SIfA), School of Physics, University of Sydney, NSW 2006, Australia} 
\affiliation{Stellar Astrophysics Centre, Department of Physics \& Astronomy, Aarhus University, Ny Munkegade 120, DK-8000 Aarhus~C, Denmark}

\author[0000-0001-6763-6562]{Roland Vanderspek}
\affiliation{Kavli Institute for Astrophysics and Space Research, Massachusetts Institute of Technology, Cambridge, MA 02139, USA}

\author{Geoff Wallace}
\affiliation{Space Telescope Science Institute, 3700 San Martin Drive, Baltimore, MD 21218, USA}

\author[0000-0002-4265-047X]{Joshua N.\ Winn}
\affiliation{Department of Astrophysical Sciences, Princeton University, Princeton, NJ 08544}

\begin{abstract}
We describe the catalogs assembled and the algorithms used to populate the revised TESS Input Catalog (TIC), 
based on the incorporation of the {\it Gaia\/} second data release.
We also describe a revised ranking system for prioritizing stars for 2-minute cadence observations, and assemble a revised Candidate Target List (CTL) using that ranking. 
The TIC is available on the Mikulski Archive for Space Telescopes (MAST) server, and an enhanced CTL is available through the Filtergraph data visualization portal system at the URL \url{http://filtergraph.vanderbilt.edu/tess_ctl}.

%
\end{abstract}

\section{Introduction\label{sec:intro}}
The TESS Input Catalog (TIC) is a comprehensive collection of sources on the sky, for use by the TESS mission to select target stars to observe, and to provide stellar parameters useful for the evaluation of transit signals. The TIC is intended to enable the selection of optimal targets for the planet transit search, to enable calculation of flux contamination in the TESS aperture for each target, and to provide reliable stellar radii for calculating planetary radii, which in turn determines the targets that will receive mission-supported photometric and spectroscopic follow-up. The TIC is also essential for the community to select targets through the Guest Investigator program.

The area of the sky projected onto each TESS pixel is large (21$\times$21 arcseconds) and the point spread function is typically 1--2 pixels in radius (depending on stellar brightness and position in the focal plane). Consequently, the photometric aperture surrounding a given TESS target may include flux from multiple objects. Therefore, it is important that the TIC contain every optically luminous, persistent, non-moving object in the sky, down to the limits of available wide-field photometric point source catalogs. 

An initial version of the TIC for use in the first year of TESS observations was delivered shortly before TESS launch in early 2018, and is described in detail by \citet{Stassun:2018}.
It had been intended from the start of planning for the TESS mission \citep{Ricker:2015} that the $\sim$1~billion point sources with parallaxes and proper motions expected from the {\it Gaia\/} mission \citep{Gaia:2018} would provide an ideal base catalog for the TIC. Unfortunately, the final data release schedule for {\it Gaia\/} only allowed the first data release (DR1) to be available prior to TESS launch. Consequently, the initial version of the TIC included parallaxes for only the $\sim$2~million bright stars in the {\it Tycho-Gaia\/} Astrometric Solution \citep[TGAS;][]{Gaia:2016}. By necessity, then, the majority of the $\sim$470~million stars in the TIC had their properties calculated from broadband photometry (and spectroscopy in a tiny minority of cases), on the basis of a complex set of logic rules, algorithms, and empirical relations, several of which had to be customized for the TESS bandpass. 

Importantly, as a result of the small number of stars with measured parallaxes, it was not possible to calculate radii accurately for the vast majority of the stars in the TIC. Therefore, it was necessary to use a proper-motion based criterion to screen out evolved stars for the Candidate Target List (CTL), from which the $\sim$200,000 targets for 2-min cadence transit search are selected, according to the TESS mission requirements (see Section~\ref{sec:ctl}). Because the proper-motion based method is not able to distinguish subgiants from dwarfs, the CTL inevitably included a large number of subgiants; we estimated that as many as $\sim$50\% of the stars in the CTL were subgiants \citep[see][for a detailed discussion]{Stassun:2018}. 
Finally, in order to ensure inclusion of known, high-value targets, the TIC and CTL were manually populated by a set of specially curated lists, including 
a Bright Star list, 
a Cool Dwarf list, a list of Known Planet Hosts, and a list of Hot Subdwarfs (see Appendix~\ref{sec:lists} for a detailed discussion).

Shortly after the TESS launch, {\it Gaia\/} delivered its second data release \citep[DR2;][]{Gaia:2018}, which includes parallaxes as well as estimated stellar properties for 1.3 billion stars. In addition to enabling a more direct determination of the stellar radii, and therefore a more optimized selection of transit targets for the CTL, the availability of uniform photometry via the three {\it Gaia\/} bandpasses ($G$, $G_{BP}$, $G_{RP}$) greatly simplifies the process of calculating various stellar properties via a smaller, consolidated set of algorithms and empirical relations. 

The purpose of this paper is to describe the updated TIC and CTL. 
Section~\ref{sec:tic} describes the construction of the TIC and the algorithms used to calculate various stellar quantities. 
Section~\ref{sec:ctl} describes the construction of the CTL, the additional algorithms used for parameters unique to the CTL, and in particular the prioritization scheme for selecting targets for 2-minute cadence observations. Finally, Section~\ref{sec:discussion} provides a summary of the contents of the TIC and CTL.

The TIC and CTL are also accompanied by official release notes, which are provided on the MAST server. Public access to the TIC is also provided via the MAST server, and access to an enhanced CTL is provided via the Filtergraph data visualization service at the URL \url{https://filtergraph.vanderbilt.edu/tess_ctl}.

\section{The TESS Input Catalog (TIC)}\label{sec:tic}

In this section we detail the 
algorithms, relations, and rules adopted for populating the TIC. The TIC includes a number of columns, each with a specified format and a permitted range of values. These are summarized by \citet{Stassun:2018}. The provenance flags associated with various TIC quantities are listed in Appendix~\ref{sec:flags}. 
{\it It is important to understand that, as described below, the TIC deliberately includes both point sources (stars) and extended sources (e.g., galaxies); positional searches of the TIC will in general return some extended sources as well as stars. For more detailed discussion about extended sources in the TIC, see \citet{Stassun:2018}. These can be separated by use of the {\tt objtype} flag (see Appendix~\ref{sec:flags}).} 
Finally, a number of specially curated lists are summarized in Appendix~\ref{sec:lists}. 

\subsection{Assembly of the TIC}

\begin{figure}[!ht]
    \centering
    \includegraphics[width=0.975\linewidth]{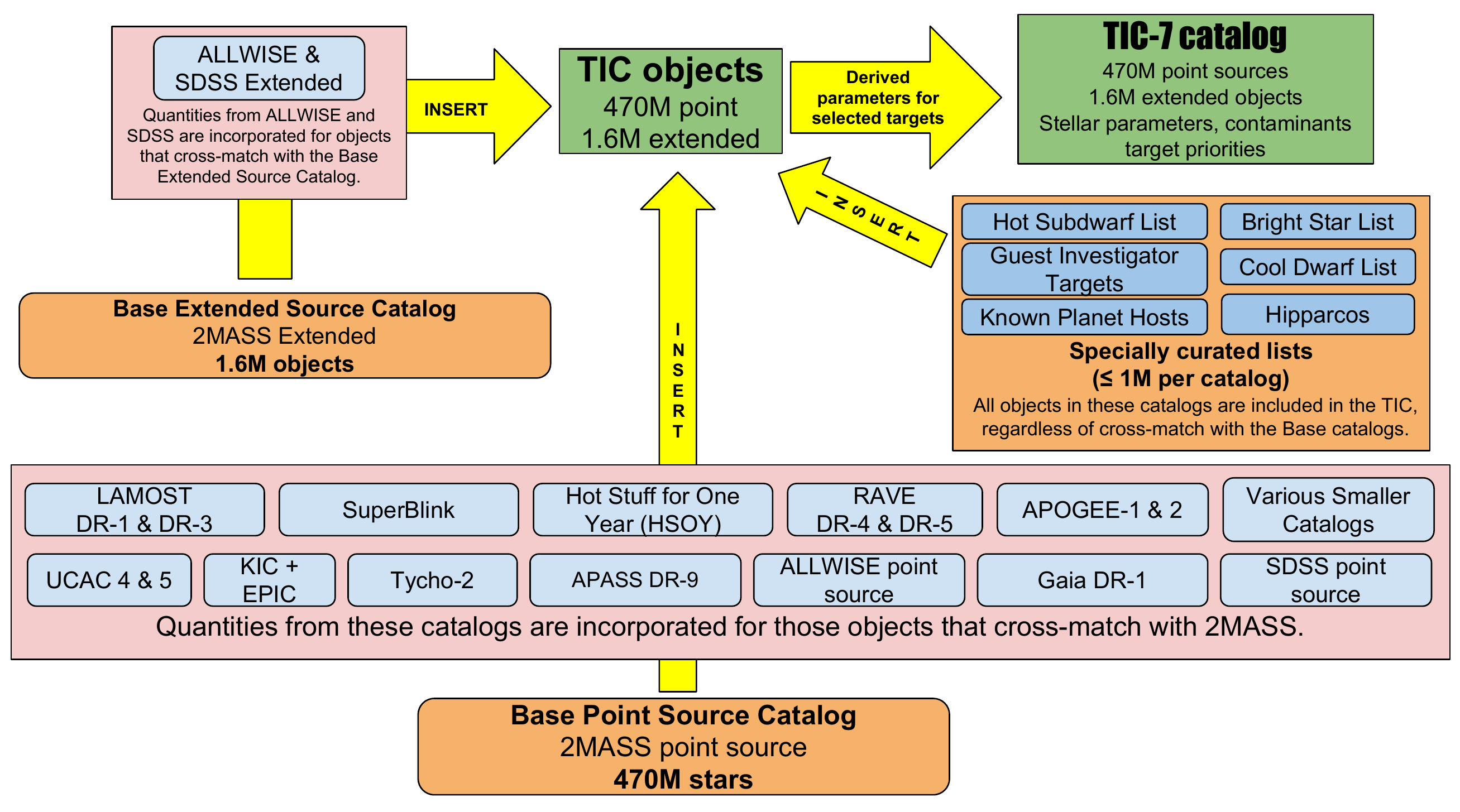}
    \caption{A visual representation of the assembly of TICv7 as produced for the first year of the TESS mission \citep[reproduced from Fig.~1 of][]{Stassun:2018}.}
    \label{fig:tic_overview_old}
\end{figure}

\begin{figure}[!ht]
    \centering
    \includegraphics[width=0.975\linewidth]{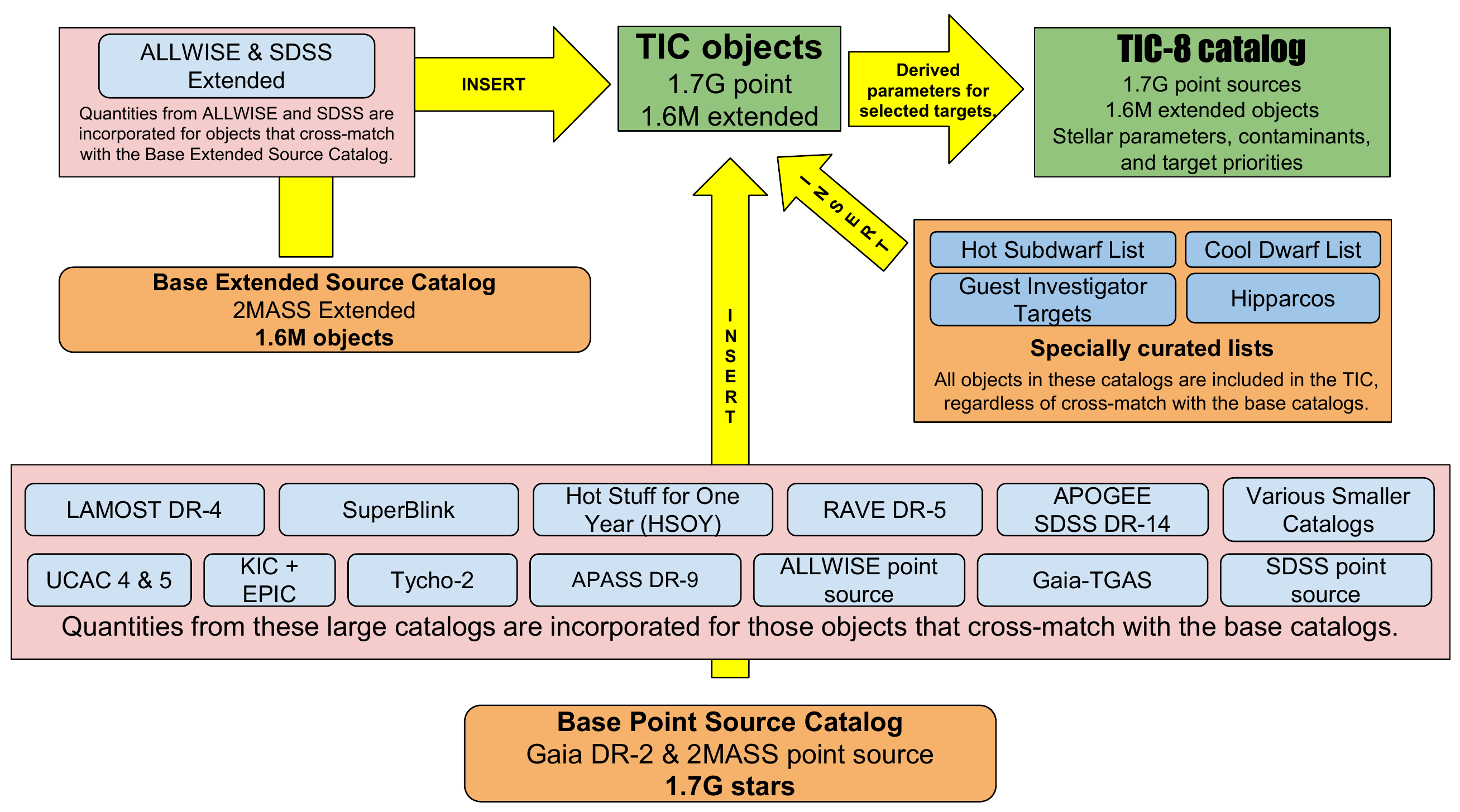}
    \caption{A visual representation of the assembly of TICv8.} 
    \label{fig:tic_overview}
\end{figure}

For the TIC that was produced for the first year of the TESS mission \citep{Stassun:2018}, we adopted the 2MASS catalog as the base point-source catalog, and referenced all other data to the 2MASS point source. As shown in Figure~\ref{fig:tic_overview_old}, reproduced from \citet{Stassun:2018} for convenience, this included a large number of catalogs containing spectroscopic quantities, photometry in the various passbands that we utilized for estimating various stellar properties, proper motions, and a number of other ancillary data. Owing to the fact that the {\it Gaia\/} DR2 catalog was not yet available at the time of TESS launch, the TIC that was delivered at that time only included the {\it Gaia\/} DR1 parallaxes, again cross-referenced to the 2MASS base catalog. 
Now we have rebuilt the TIC with {\it Gaia\/} DR2 as the base. An updated visual overview, analogous to Fig.~\ref{fig:tic_overview_old}, is shown in Figure~\ref{fig:tic_overview} and described in detail in the following subsections.

\subsection{Point Sources}

The base point-source catalog for the TIC is {\it Gaia\/} DR2. In order to preserve continuity and provenance with the previous version of the TIC, which was based on the 2MASS catalog, we first translated all previous TIC sources to the new TIC catalog using the association between {\it Gaia\/} and 2MASS that is provided within {\it Gaia\/} DR2 itself.

TIC coordinates and their uncertainties have been propagated to epoch 2000 due to mission requirements. The error propagation leads to much larger uncertainties than those native to the nominal Gaia DR2 positions. Especially for Gaia DR2 stars, users should not try to propagate forward the TIC coordinates using the proper motions listed. Instead, users should use the original Gaia DR2 positions, proper motions, and corresponding errors for propagation. We provide the original RA and Dec with errors as given in the source catalog (Gaia DR2, 2MASS etc) in additional columns on MAST and \textit{Filtergraph}. 


Due to the improved angular resolution and depth of the {\it Gaia\/} DR2 catalog relative to 2MASS, there were a large number of cases where a single 2MASS source turned out to be associated with two {\it Gaia\/} sources. In these cases, we retain the association of the one 2MASS identifier with both {\it Gaia\/} sources, however we set the $JHK_S$ magnitudes of both sources to null because there was no definitive means for splitting the reported 2MASS flux among the two {\it Gaia\/} sources. While we have not done so here for the sake of catalog purity, we note that in principle it is possible to estimate the 2MASS $JHK_S$ magnitudes for the two sources from the {\it Gaia\/} reported $G G_{BP} G_{RP}$ fluxes and the relations provided by \citet{Evans:2018}. For the purposes of the TIC, we require only the {\it Gaia\/} reported $G G_{BP} G_{RP}$ magnitudes, as described below, to which we applied the corrections for bright stars ($G < 6$) as reported by \citet{Evans:2018}.
There were also $\sim$33~million cases of 2MASS sources that had no {\it Gaia\/} counterpart. 
We expect that many of these stars are 2MASS artifacts around bright stars, but we did not identify a straightforward way to identify them consistently, and therefore have left them unaltered for TICv8. 

While we calculate the stellar properties for most TIC stars from {\it Gaia\/} magnitudes via relations discussed below, where possible we adopt measured spectroscopic parameters. Following the conventions of the initial TIC, we selected effective temperatures (\teff) and metallicities ([Fe/H]) when available, from the following catalogs and in the order of preference shown in Table~\ref{tab:spectra_cats}. Users are cautioned that surface gravities (\logg) are always \textit{calculated} in TICv8 using the TICv8 reported mass and radius; \logg\ is not adopted from spectroscopic catalogs even when available so as to ensure internal consistency of \logg\ with mass and radius. 
Note also that while metallicities are reported when available from spectroscopic catalogs, this is for convenience only and we do not use metallicity in any relations or derived quantities.

\begin{center}
\vspace{-0.1in}
\begin{longtable}[c]{|c|c|c|c|c|}
 \hline
 Name & Data Release & Approximate Num. of Stars & Priority & Reference \\
 \hline
 SPOCS & & 1.6~k & 1 & \citet{Brewer:2016}\\
 PASTEL & & 93~k & 2 & \citet{Soubiran:2016}\\
 {\it Gaia}-ESO & DR-3 & 29~k & 3 & \citet{Gilmore:2012}\\
 {\it TESS}-HERMES & DR-1 & 25~k & 4 & \citet{Sharma:2018}\\
 GALAH & DR-2 & 340~k & 5 & \citet{Buder:2018}\\
 APOGEE-2 & DR-14 & 277~k & 6 & \citet{Abolfathi:2017}\\
 LAMOST & DR-4 & 2.9~M & 7 & \citet{Luo:2015}\\
 RAVE & DR-5 & 484~k & 8 & \citet{Kunder:2017}\\
 Geneva-Copenhagen & DR-3 & 16~k & 9 & \citet{Holmberg:2009}\\
 \hline
 \caption{Spectroscopic Catalogs in the TIC.}
 \label{tab:spectra_cats}
\end{longtable}
\vspace{-0.5in}
\end{center}


\subsection{Algorithms for calculated stellar parameters}\label{subsec:tic_algorithms}

In this section, we describe the algorithmic procedures we adopted for calculating various stellar parameters. We begin with the procedure for calculating an apparent magnitude in the TESS bandpass, \tmag, as this is the most basic quantity required of any object in the TIC. Since many of the subsequent empirical relations that we adopt depend on the effective temperature (\teff), we next describe the procedures for determining \teff. Briefly, we prefer spectroscopic \teff\ if available and if the reported error is less than 300~K, otherwise 
we calculate \teff\ from photometric colors via empirical relations that we describe. These photometric colors must first be corrected for reddening, thus we first also describe our photometric dereddening procedures. 


We next apply cuts based on the {\it Gaia\/} DR2 quality flags on astrometry and photometry \citep[see][equations 1 and 2]{Arenou:2018}, such that objects failing these quality criteria do not have any other stellar parameters computed. 
In these cases, if stellar parameters had been computed for the CTL in TICv7, then we adopt those parameters again here. 

Note that wherever our relations involve the {\it Gaia\/} DR2 parallax, we utilize the Bayesian distance estimate from \citet{Bailer-Jones:2018}. This both provides a proper posterior estimate of the (generally asymmetric) errors in the distance, and assures that the distance estimate is non-negative, since in some cases the native {\it Gaia\/} DR2 parallax can be negative. 
Finally, we defer a detailed discussion of the procedure for calculating parameter uncertainties to Section~\ref{sec:ctl_algorithms}, where we describe a Monte Carlo based approach that we apply to objects in the CTL. Here we simply note that, for the TIC, which requires symmetric error bars to be reported, we take the arithmetic mean of the asymmetric error bars determined for CTL objects. In some cases the resulting symmetrized error can be larger than the quantity itself; these cases should be regarded with caution.

\subsubsection{TESS Magnitude}\label{subsubsec:tmag}


The most basic quantity required for every TIC object, aside from its position, is its apparent magnitude in the TESS bandpass, 
which we represent as \tmag. 
As we did in TICv7, we derived a relation based on the PHOENIX model atmospheres \citep{Husser:2016}. We adopted the most up-to-date TESS passband available (R.\ Vanderspek, private communication) and the 
{\it Gaia\/} passbands for $G$, \gbp, and \grp\ 
from \citet{Gaia:2018}. 
Note that we did not apply the small corrections to the {\it Gaia\/} bandpasses from \citet{Maiz:2018} as these were not available prior to our construction of the base TIC. 

\begin{figure}[!ht]
    \centering
    \includegraphics[width=0.37\linewidth,angle=270,trim=70 0 50 0,clip]{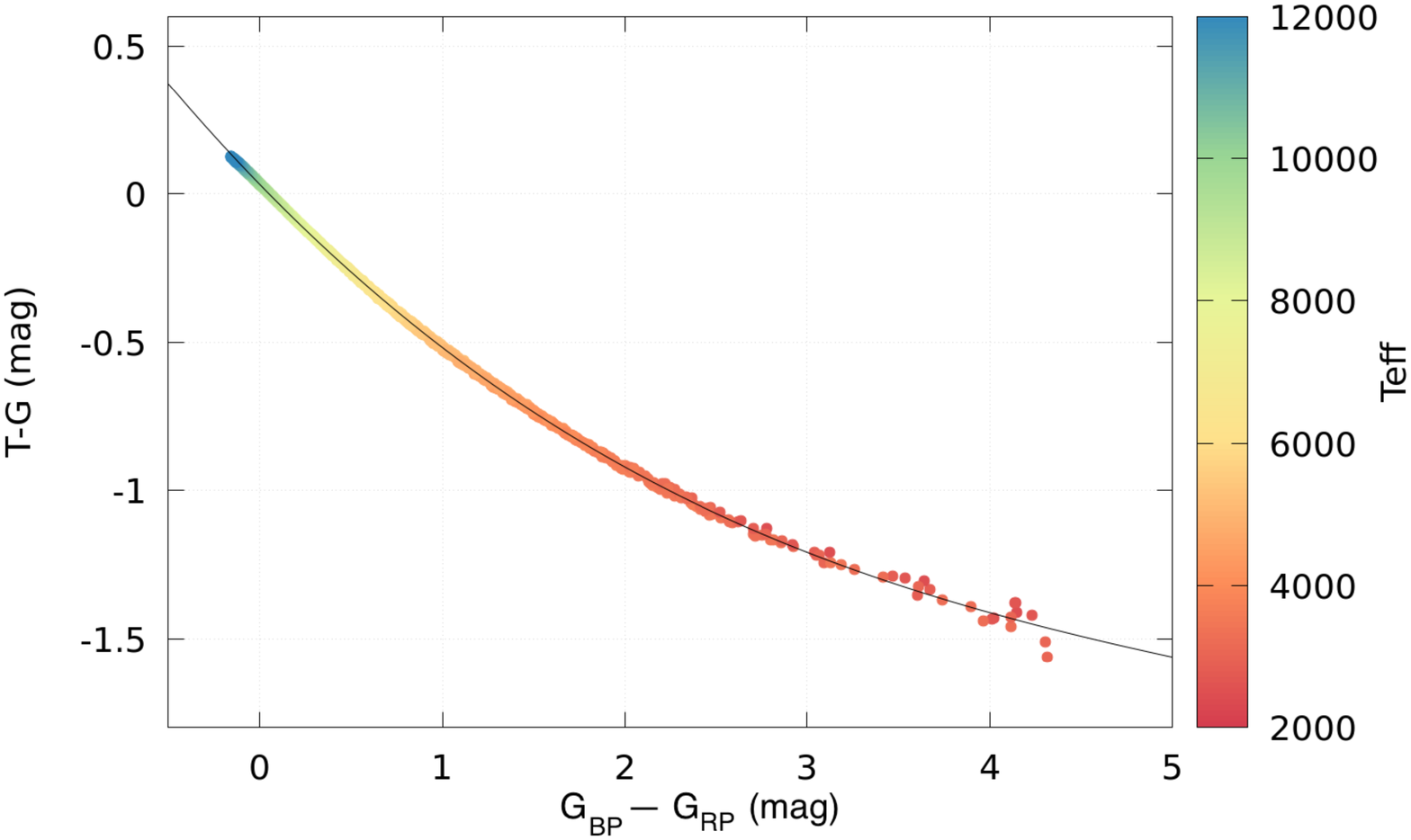}
    \includegraphics[width=0.37\linewidth,angle=270,trim=70 0 50 0,clip]{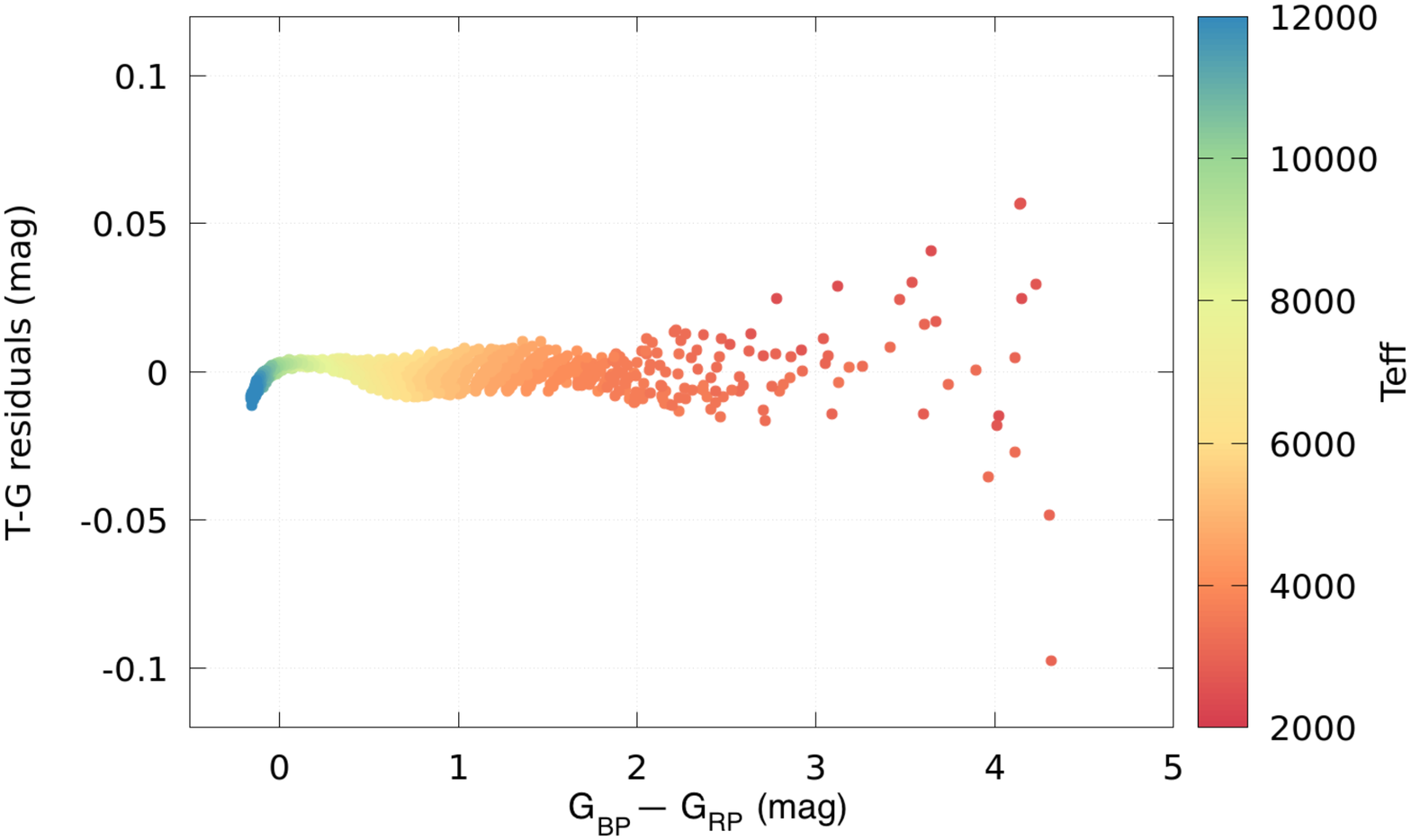}
    \caption{Derivation of our TESS magnitude relation. A polynomial fit to synthetic colors from PHOENIX models is shown at the top (equation given in the text), and residuals at the bottom.}
    \label{fig:T_relations}
\end{figure}

The relation that we adopt, 
\begin{equation}
T = G -0.00522555 (\gbp - \grp)^3 + 0.0891337 (\gbp - \grp)^2 - 0.633923 (\gbp - \grp) + 0.0324473 \, ,
\end{equation}
is valid for dwarfs, subgiants, and giants of any metallicity, and the formal scatter is 0.006~mag. The fit and residuals are shown in Figure~\ref{fig:T_relations}. 
Strictly speaking, this relation is valid for 
$-0.2 < \gbp - \grp < 3.5$ but we extrapolate it to $-1.0 < \gbp - \grp < 6.0$ because by NASA requirement every star in the TIC must have a \tmag\ magnitude. Even though the relation degrades considerably for M~dwarfs $(\gbp - \grp > 2)$, we consider it to be the best available estimator from the $G$ band. Most importantly, the refined \tmag\ magnitudes provided in the specially curated Cool Dwarf list override the magnitudes computed by the relation above.   

\begin{figure}[!ht]
    \centering
    \includegraphics[width=0.37\linewidth,angle=270,trim=60 0 0 0,clip]{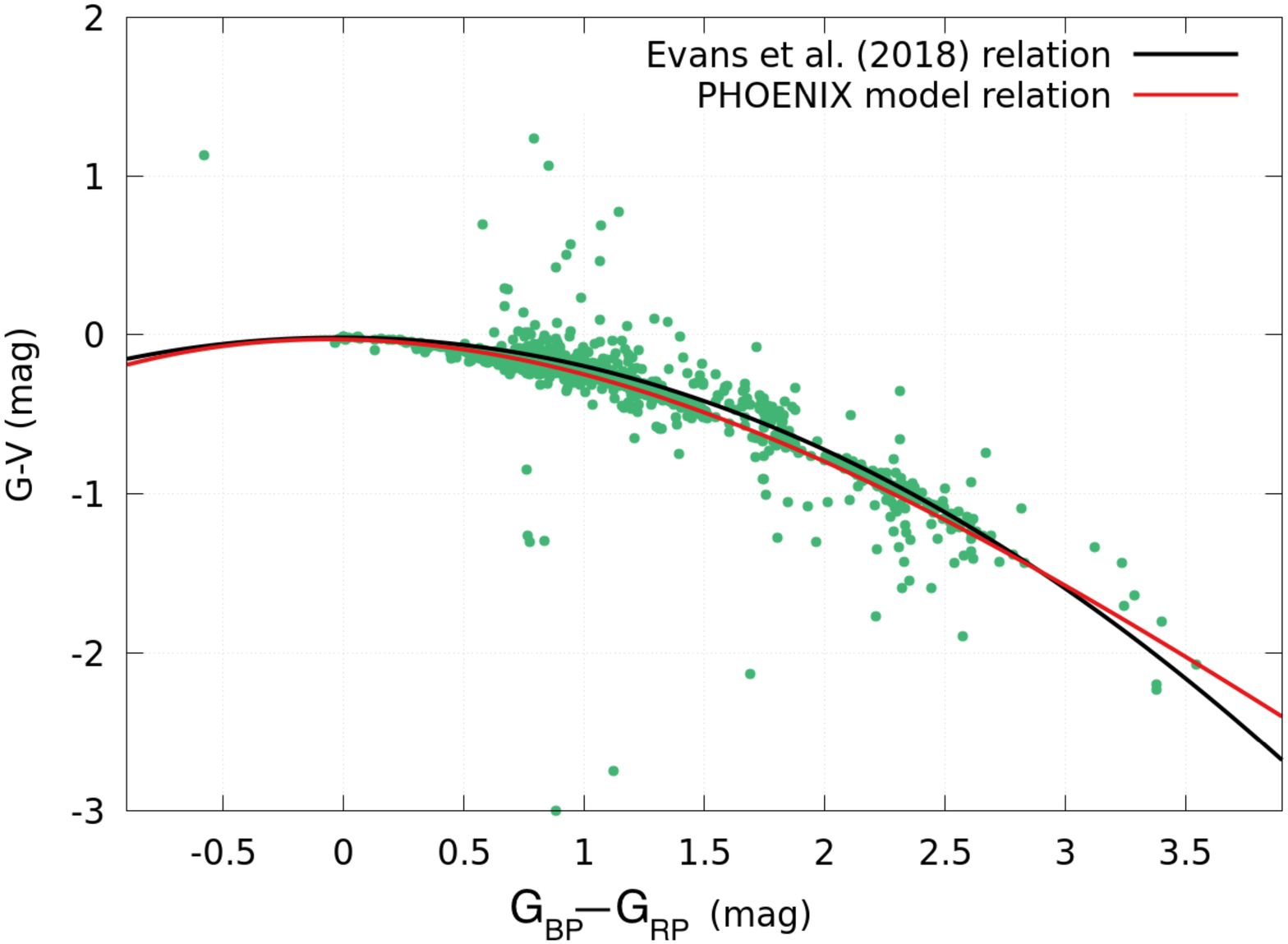}
    \includegraphics[width=0.37\linewidth,angle=270,trim=60 0 0 0,clip]{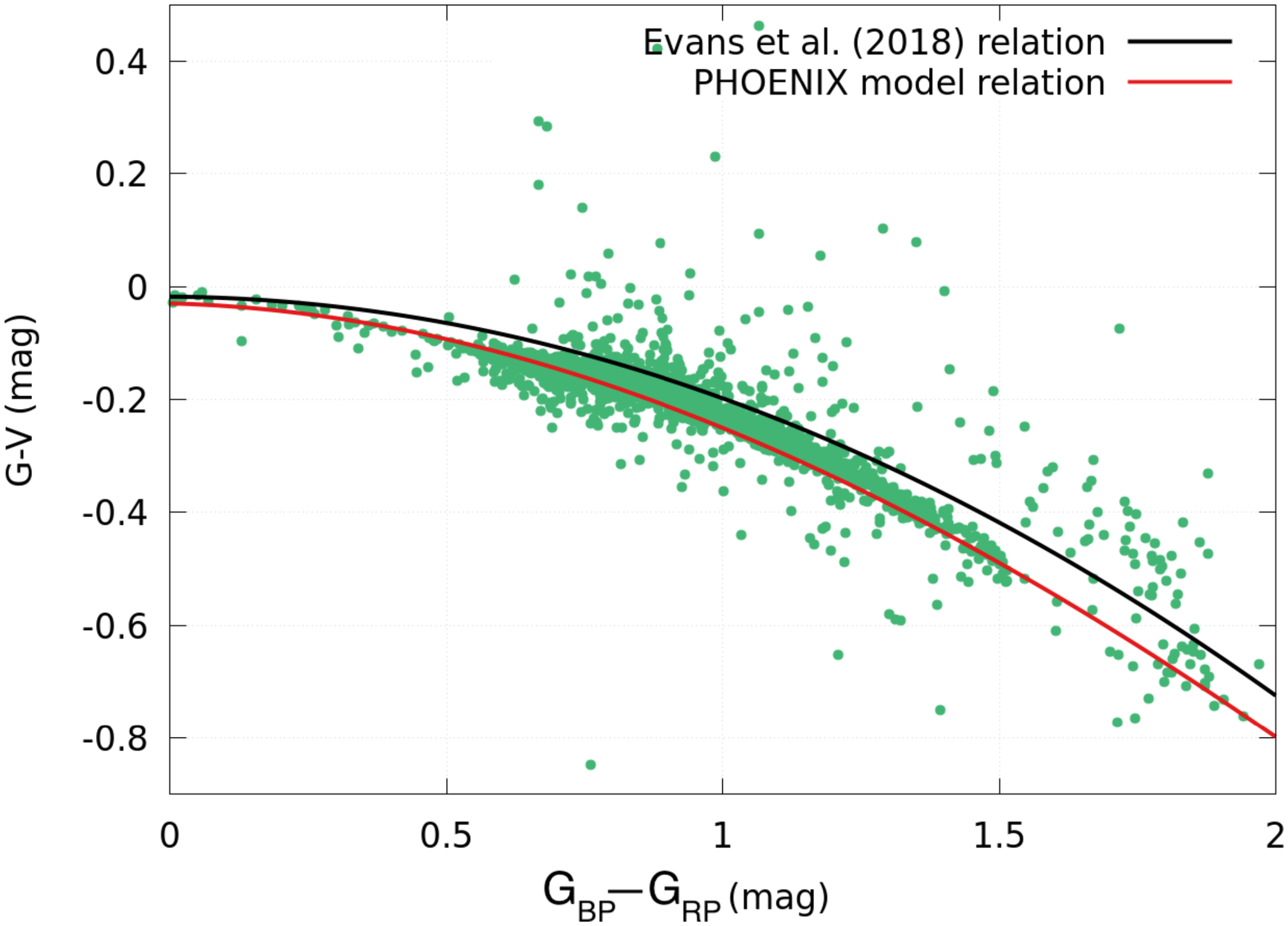}
    \caption{The top panel presents a comparison of our synthetic $G-V$ colors as a function of $\gbp-\grp$ (polynomial red line fit) with a similar relation by \cite{Evans:2018} shown in black. The bottom panel shows a closeup of the solar-color region. The points represent actual measurements for a set of stars from TICv7.}
    \label{fig:GV_comp}
\end{figure}

Of course, the use of stellar atmosphere models introduces some systematic error, so the true errors in the predicted \tmag\ are likely to be larger than 0.006~mag. To estimate this, we used the same atmosphere models to derive a relation between $V$ magnitude and the magnitudes in the {\it Gaia\/} passbands, and compared our relation with the empirical relation reported by \citet{Evans:2018}. Their empirical relation is based on real $V$ magnitudes for stars from various catalogs and from the measured $G$ magnitudes for the same stars from {\it Gaia\/} DR2. Figure~\ref{fig:GV_comp} compares the two relations. 
The comparison here is based on the set of stars from TICv7 that had spectroscopic \teff, thus they are not the same set of stars used by \citet{Evans:2018}; however, this should not affect our conclusions significantly. 
A slight difference is evident between our model fit and the \citet{Evans:2018} empirical fit, which is to be expected. 
However, the largest difference between the two relations is only $\sim$0.1~mag, providing confidence in our adopted relation and suggesting that the true uncertainties in our derived \tmag\ are likely to be at most $\sim$0.1~mag in most cases.

The TESS magnitude relation above is strictly valid for unreddened stars. Because the measured magnitudes and colors from {\it Gaia\/} are typically affected by extinction, we first de-redden the colors and correct the $G$ magnitudes for extinction, then apply the relation to obtain a de-reddened $T$ magnitude, and finally add extinction back into $T$ to obtain an apparent magnitude. The extinction coefficients required in each band are provided below in Section~\ref{sec:extinction}.

Finally, we have developed simple relations 
for stars that either have colors beyond the formal validity limits of the above relation or that do not have fluxes reported for all three {\it Gaia\/} passbands. 
For stars that are bluer or redder than the limits of the above relation, we simply extrapolate the same polynomial, however to be conservative we increase the formal errors by 0.1~mag (added in quadrature). 
For stars with no valid $\gbp-\grp$ colors, but which have a valid $G$ magnitude, we use the following simple offset:
\begin{equation}
T = G - 0.430 \, .
\end{equation}
This offset is the value that specifically corresponds to a star like the Sun, with a color of $\gbp-\grp = 0.82$ based on the PHOENIX stellar atmosphere models. As a very conservative error, we assign 0.6~mag, which should be valid for all but the reddest M dwarfs, which are in any case dealt with via the specially curated Cool Dwarf list.

\subsubsection{V magnitude}
For completeness and for maximum usability, we compute the apparent $V$ magnitude for TIC stars that do not possess a measured $V$ from TICv7 but which possess {\it Gaia\/} photometry, as follows, using the relations provided by the {\it Gaia\/} team \citep{Evans:2018}: 
\begin{equation}
V = G + 0.01760 + 0.006860 (\gbp-\grp) + 0.1732 (\gbp-\grp)^2 \, , 
\end{equation}
which has a reported scatter of about 0.046~mag.

\subsubsection{Extinction and Dereddening\label{sec:dereddening}}
\label{sec:extinction}


Because we estimate stellar \teff\ principally from an empirical color relation involving the {\it Gaia\/} $\gbp-\grp$ color (see Section~\ref{subsubsec:teff}), which is susceptible to reddening effects, it is necessary to first apply a dereddening correction, as we now describe. 

When creating TICv7, we were limited by the available dust maps to estimates of the reddening along the full line of sight through the Galaxy in any particular direction, and also were unable to estimate reddening within about 15 degrees of the Galactic plane. Here, we adopt the newly released three-dimensional, empirical, nearly all-sky dust maps from Pan-STARRS \citep{Green:2018}, which provides an ability to estimate the reddening on a star-by-star basis according to the star's position in three dimensions (coordinates on the plane of the sky together with the distance). For the region of the sky not covered by Pan-STARRS (Declination below $-30\arcdeg$), we continue to use the \citet{Schlegel:1998} map, now with an adjustment to the total line-of-sight extinction for distance (from {\it Gaia}), assuming a standard exponential model for the disk with a scale height of 125~pc \citep[see, e.g.][]{Bonifacio:2000}. In both cases we apply a recalibration coefficient of 0.884 to the $E(B-V)$ values, as prescribed by \citet{Schlafly:2011}.

\begin{figure}[!ht]
    \centering
    \includegraphics[width=0.49\linewidth]{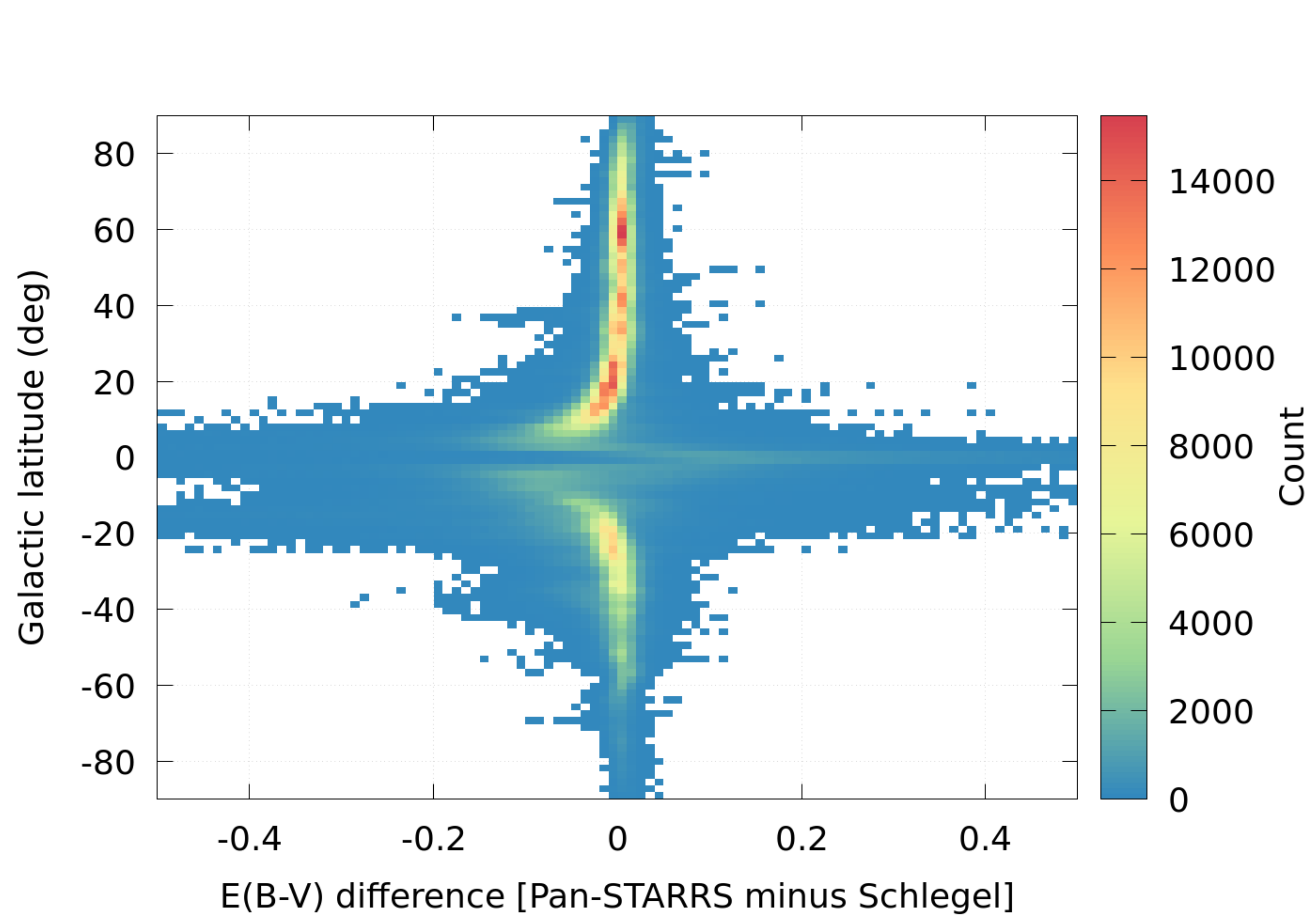}
    \includegraphics[width=0.49\linewidth]{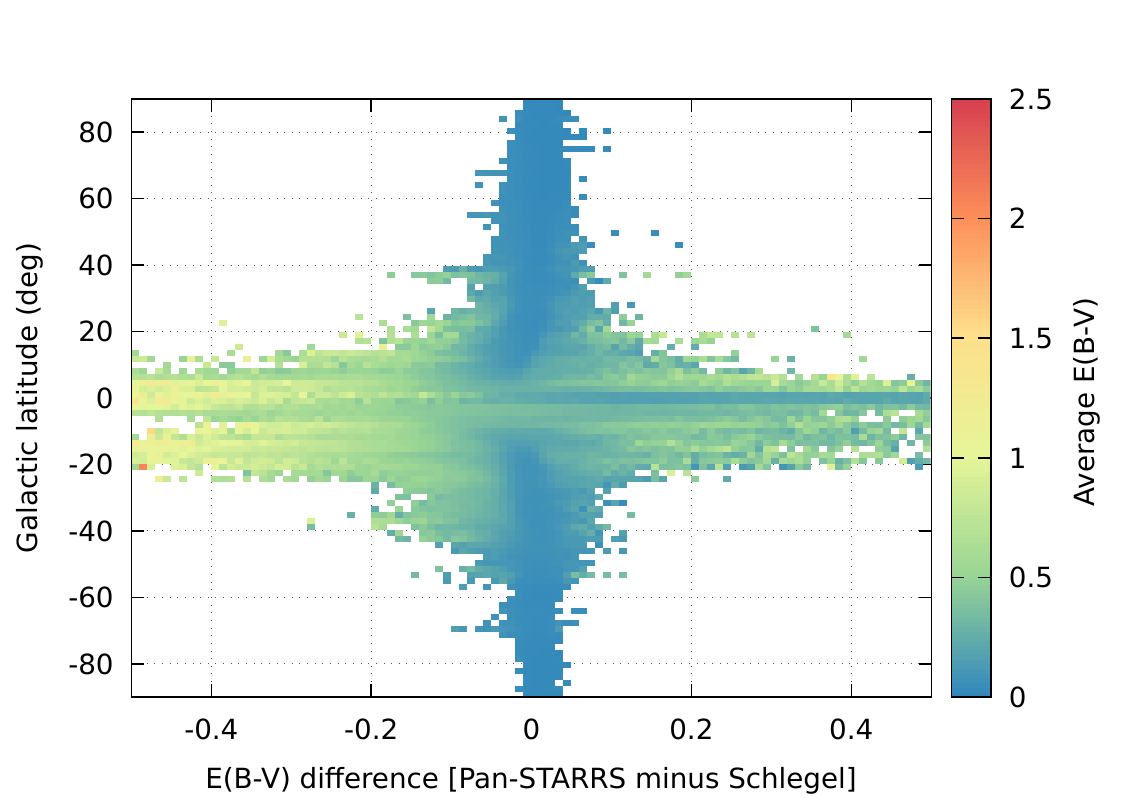}
    \caption{(Left:) Comparison between the $E(B-V)$ reddening values returned by the Pan-STARRS and Schlegel extinction maps, as a function of Galactic latitude. (Right:) Same, but color-coded by average of $E(B-V)$.}
        \label{fig:dered_ebv_comp}
\end{figure}

As an initial sanity check on the Pan-STARRS reddening estimates, we compared the $E(B-V)$ reddening values reported by the Pan-STARRS map against that estimated from the dust maps that we used in TICv7 \citep{Schlegel:1998}. As shown in Figure~\ref{fig:dered_ebv_comp}, the agreement is in general quite good, with the $E(B-V)$ agreeing to within $\sim$0.05~mag for Galactic latitudes $|b| > 15$~deg. 

We require a relation to convert the $E(B-V)$ values that are provided by the reddening maps into $E(\gbp-\grp)$ and $A_G$ for dereddening the {\it Gaia\/} $\gbp-\grp$ and $G$ observed colors. 
As the most common type of star in the TIC is similar to the Sun, we used a synthetic solar-like spectrum as the source (\teff = 5800~K, \logg = 4.5, [Fe/H] = 0.0, the closest PHOENIX model spectrum to the Sun) to compute the effective wavelengths for the various passbands following Eq.~18 of \citet{Casagrande:2014}.
We then used these mean wavelengths with the \citet{Cardelli:1989} extinction law, which yields $E(\gbp-\grp) = 1.31 E(B-V)$ and $A_G = 2.72 E(B-V)$. 
We also used this procedure to determine the relation of $E(B-V)$ to $A_T$, the extinction in the TESS bandpass, for use in calculating the \tmag\ magnitudes (see Section~\ref{subsubsec:tmag}), which gives:
$A(T) = 2.06 E(B-V)$. 

\begin{figure}[!ht]
    \centering
    \includegraphics[width=0.7\linewidth]{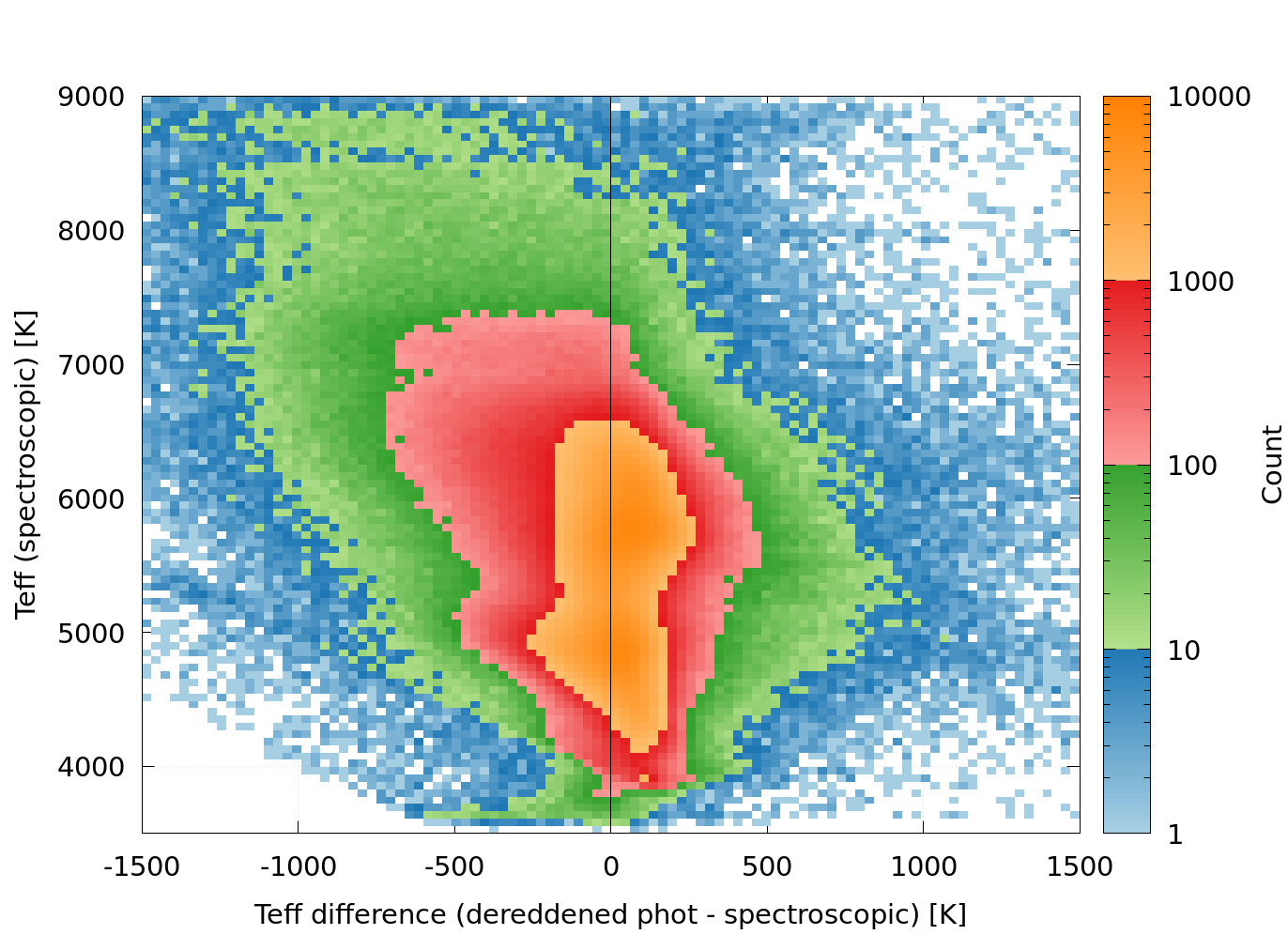}
    \includegraphics[width=0.495\linewidth]{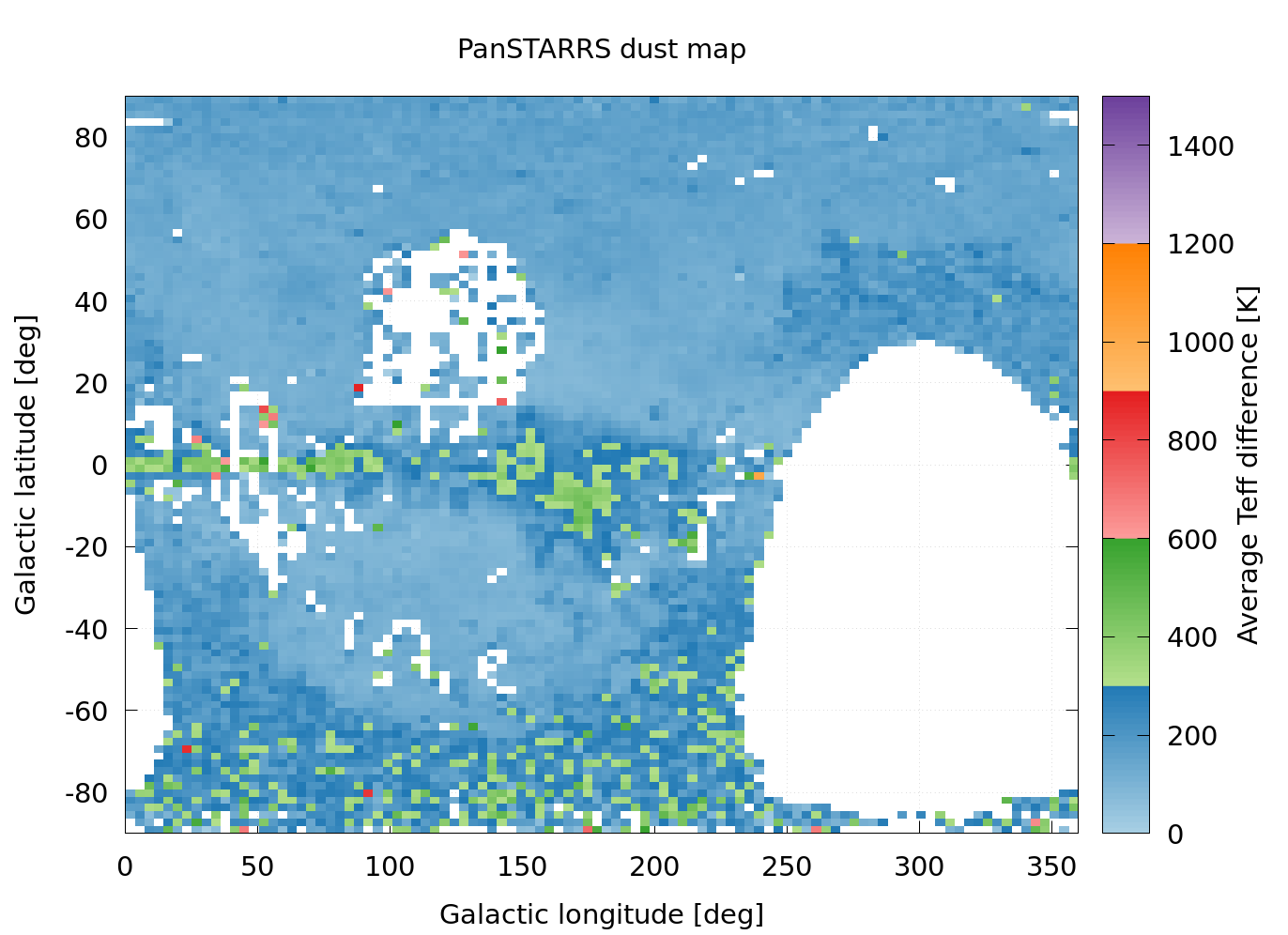}
    \includegraphics[width=0.495\linewidth]{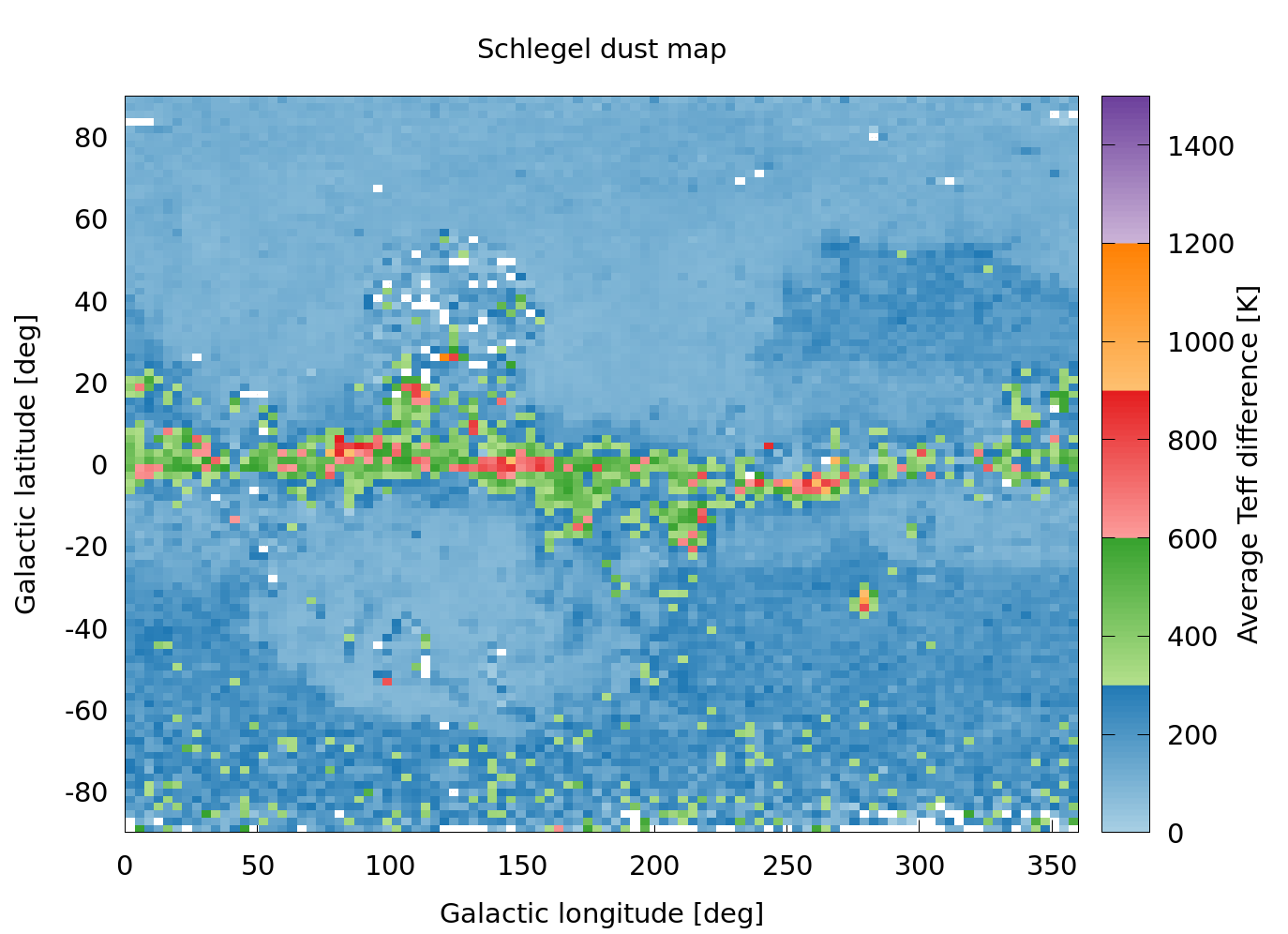}
    \caption{Comparison of \teff\ from dereddened photometric colors versus spectroscopic. The median \teff\ difference is $\approx$20~K, and the 95th percentile range is $-410$~K to $+220$~K. }
    \label{fig:teff_diff_map}
\end{figure}

Applying the estimated reddening to the determination of \teff\ via the photometric colors has the effect generally of making apparently cool stars hotter, and that hotter \teff\ in turn has the effect of inferring a larger radius and mass from our empirical relations described in Section~\ref{sec:mass_radius}. Therefore it is important to assess the quality of the dereddened \teff. 
Figure~\ref{fig:teff_diff_map} (top) shows a comparison of the \teff\ obtained from dereddened colors versus \teff\ measured spectroscopically for $\sim$2~million stars from TICv7 with the relevant quantities available. As with all of the spectroscopic \teff\ that we adopt in the TIC, we limited the sample to stars whose spectroscopic \teff\ have reported uncertainties less than 300~K to ensure that the spectroscopic \teff\ does not dominate the comparison of errors. The comparison overall is quite good, with a mean difference of 20~K and a 95th percentile range of $-410$~K to $+220$~K. The slight skew toward negative \teff\ differences (i.e., dereddened photometric \teff\ being slightly cooler than the spectroscopic \teff) suggests that in some cases the reddening is underestimated (not enough dereddening correction applied), however this is at the margins of the overall distribution. 
Note that both $G$ and \tmag\ are broad photometric bands, which can complicate extinction corrections. In particular, the ratio of total to selective extinction, $R_G \equiv A_G/E(B-V)$,
is a function of \teff\ as well as the overall extinction $A_V$.
The variation with $A_V$ is small (${dR_G}/{d A_V} \approx -0.03$) and can be ignored, but the variation with \teff\ is slightly larger.
This could also be a reason for the systematic trends seen in Figure~\ref{fig:teff_diff_map} (top); see Figure~A of \citet{Sanders:2018} for further details and for one way to account for this.

\begin{figure}[!ht]
    \centering
    \includegraphics[width=0.7\linewidth]{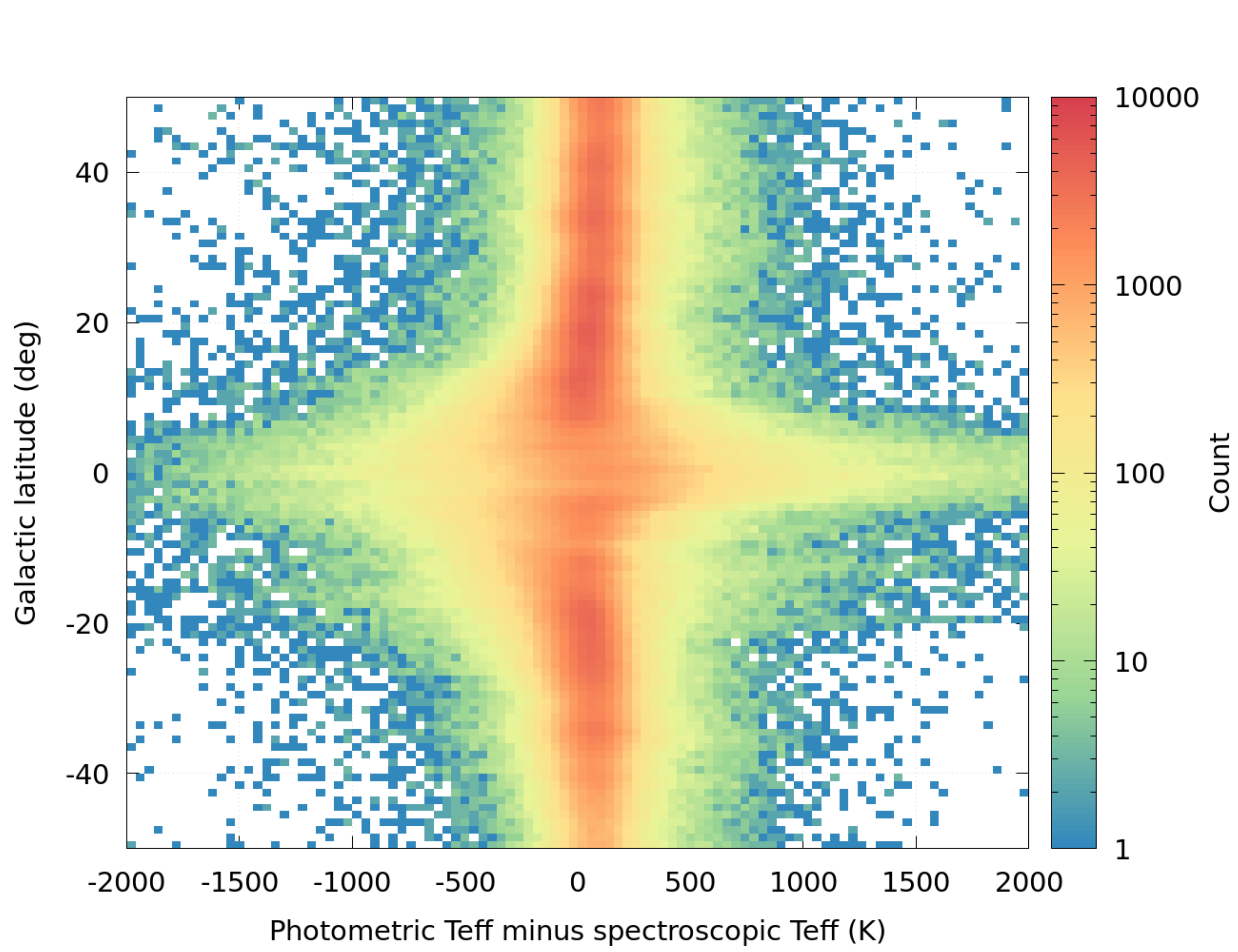}
    \caption{Comparison between photometric temperatures (based on {\it Gaia\/} colors de-reddened using the Pan-STARRS dust map) and spectroscopic temperatures for the same stars, as a function of Galactic latitude.}
    \label{fig:glat_vs_teffdiff}
\end{figure}

Also shown in Figure~\ref{fig:glat_vs_teffdiff} are the \teff\ differences as a function of Galactic coordinates, where the effect of larger errors within $\sim$10~degrees of the plane is clear. 
Note that the \citet{Schlegel:1998} dust maps have been shown to overestimate extinction for $E(B-V) > 0.15$ by a factor of about 1.4
\citep{Arce:1999,Cambresy:2005}; a correction is given in Eq.~24 of \citet{Sharma:2014}. In any event, 
we reduce the CTL priority by a factor of 0.1 for stars within 10 degrees of the Galactic plane (see Sec.~\ref{subsec:priority}).

Finally, based on the comparisons above between the Pan-STARRS and \citet{Schlegel:1998} dust maps, we find that only 1\% of the $\sim$4~million stars compared have applied $E(B-V)$ values that disagree by more than 2.5~mag. This means that a star that in reality has a very low reddening could appear with $E(B-V) > 2.5$ if the dust map at that location is such an extreme outlier. Thus, we adopt 2.5 as the maximum permissible $E(B-V)$ from the \citet{Schlegel:1998} dust map (effectively a cap on $A_V$ of $\approx$7.8) in cases where a reliable Pan-STARRS value is not available. In cases where there is not a reliable Pan-STARRS value and our adopted value from \citet{Schlegel:1998} has been capped, we report the values, but do not apply any reddening in calculating the \tmag\ magnitude, and we also do not attempt to provide any derived stellar parameters.

\subsubsection{Effective Temperature}
\label{subsubsec:teff}
\begin{figure}[!ht]
    \centering
    \includegraphics[width=0.6\linewidth]{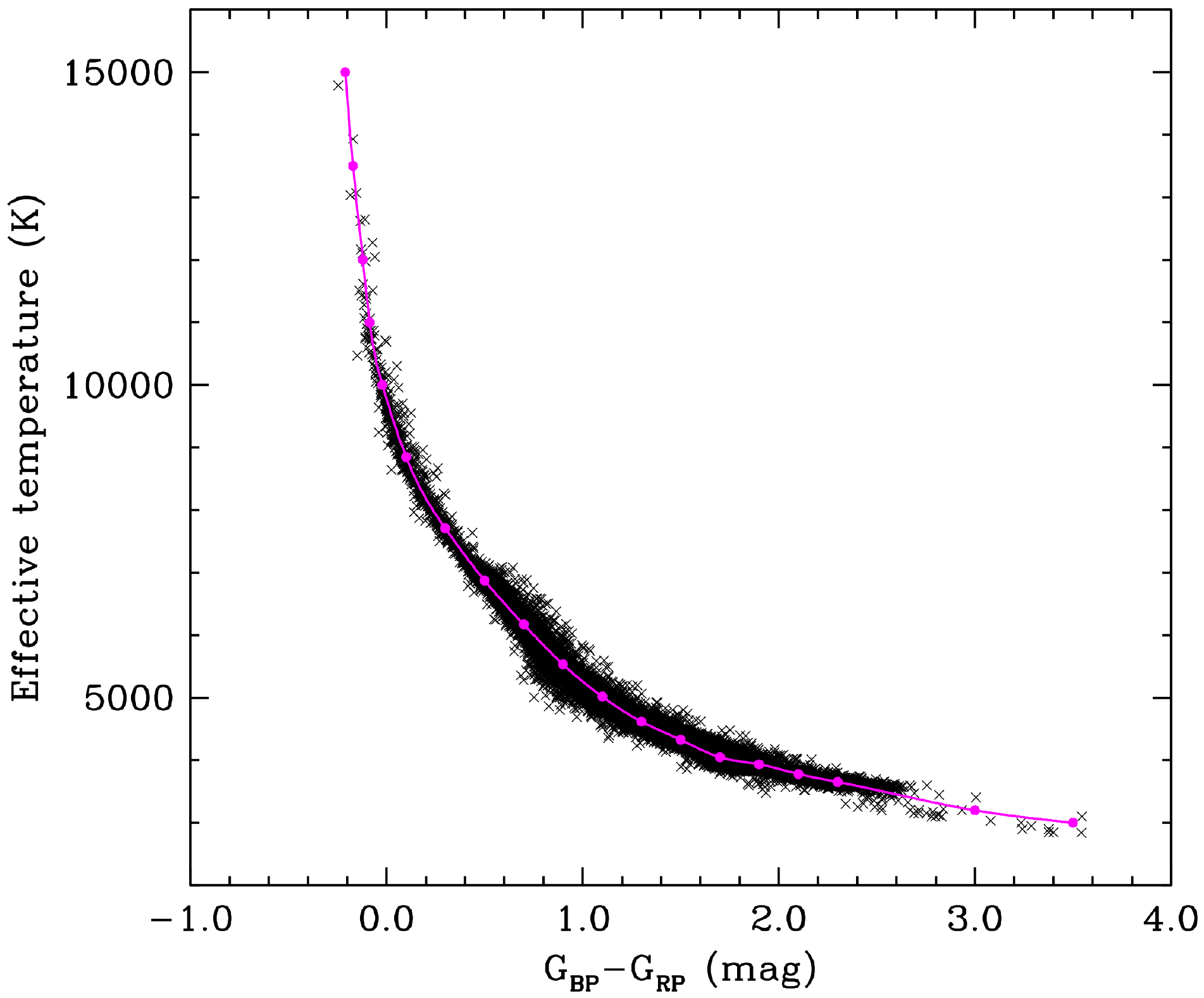}
    \caption{\teff\ as a function of the $\gbp-\grp$ color. See Table~\ref{tab:teff_index} for the nodes of the cubic spline fit.}
    \label{fig:teff_gmag_fit}
\end{figure}

We derived a new empirical relation between $T_{\rm eff}$ and the {\it Gaia\/} $\gbp-\grp$ colors, based on a set of 19,962 stars having spectroscopically determined $T_{\rm eff}$ and being within 100~pc so as to avoid reddening. 
After removing obvious outliers, we fit a spline function by eye. 
Figure~\ref{fig:teff_gmag_fit} shows the fit, with the spline nodes marked with circles. Table~\ref{tab:teff_index} lists the nodes as ($\gbp-\grp$, \teff) pairs. 

\begin{center}
\vspace{-0.1in}
\begin{longtable}[c]{|c|c|}
 \hline
 \gbp-\grp  & \teff  \\
 \hline
 -0.21 & 15000\\
-0.17 & 13500\\
-0.12 & 12000\\
-0.085 & 11000\\
-0.02 & 10000\\
0.1 & 8849.803711\\
0.3 & 7709.192383\\
0.5 & 6875.640137\\
0.7 & 6172.216309\\
0.9 & 5532.801758\\
1.1 & 5017.910156\\
1.3 & 4618.64209\\
1.5 & 4327.293457\\
1.7 & 4048.811523\\
1.9 & 3935.294434\\
2.1 & 3780.993652\\
2.3 & 3652.275635\\
3.00 &  3200\\
3.50  & 3000\\
 \hline
 \caption{Spline nodes for relation in Figure~\ref{fig:teff_gmag_fit}}
 \label{tab:teff_index}
\end{longtable}
\vspace{-0.3in}
\end{center}

The residuals of the fit (see Figure~\ref{fig:teff_diff_map}, top panel) show a near-zero mean offset, with an r.m.s.\ scatter of 122~K. 
This seems quite reasonable: given the $\sim$100~K uncertainties typical of the spectroscopic \teff, this would imply a true scatter in the photometric relation of $\sim$70~K.
The range of validity of the relation is seen in the figure, and is $\gbp-\grp = [-0.2,+3.5]$, although the predictions are likely to be less reliable for hot stars with $\gbp-\grp < 0$ because of the very steep slope of the relation, and also for M stars with $\gbp-\grp > 2$. 
As discussed below, the CTL relies mainly on the specially curated Cool Dwarf list for cool dwarf stars, and draws \teff\ estimates for those stars from that list.

This relation provides a continuous color-\teff\ relation from 3000~K to 15000~K. For stars with $\gbp-\grp$ outside
of this range of validity, the TIC reports \teff\ = {\tt Null}, unless a spectroscopic \teff\ is available or if a \teff\ is available from the CTL associated with TICv7. 
The final \teff\ errors reported in the TIC from the above polynomial relation include the 122~K scatter added in quadrature to the uncertainties from the photometric errors.

\subsubsection{Stellar Mass and Radius\label{sec:mass_radius}}

We compute the stellar radii using the {\it Gaia\/} parallaxes, which we now have for every star in the CTL, according to the standard expression from the Stefan-Boltzmann relation:  
\begin{equation}
\log (R/{\rm R}_{\sun}) = \frac{1}{5}\Big[4.74 - 5 + 5\log D - G - 10 \log (T_{\rm eff}/5772) - {\rm BC}_G \Big]
\end{equation}
where $D$ is the distance based on the {\it Gaia\/} parallax from \citet{Bailer-Jones:2018},
$G$ is the observed {\it Gaia\/} magnitude corrected for extinction ($G_{\rm obs} - A_G$), 
$T_{\rm eff}$ is the temperature from either spectroscopy or from dereddened colors, 
and BC$_G$ is the bolometric correction in the {\it Gaia\/} passband as a function of \teff\ (corrected for reddening if from colors).
To be conservative, for \teff\ from spectroscopy, we add 100~K in quadrature to the \teff\ uncertainty if the catalog reported \teff\ uncertainty is less than 100~K when computing the resulting mass and radius uncertainties.

\begin{figure}[!ht]
    \centering
    \includegraphics[width=0.5\linewidth,clip,trim=0 170 0 190]{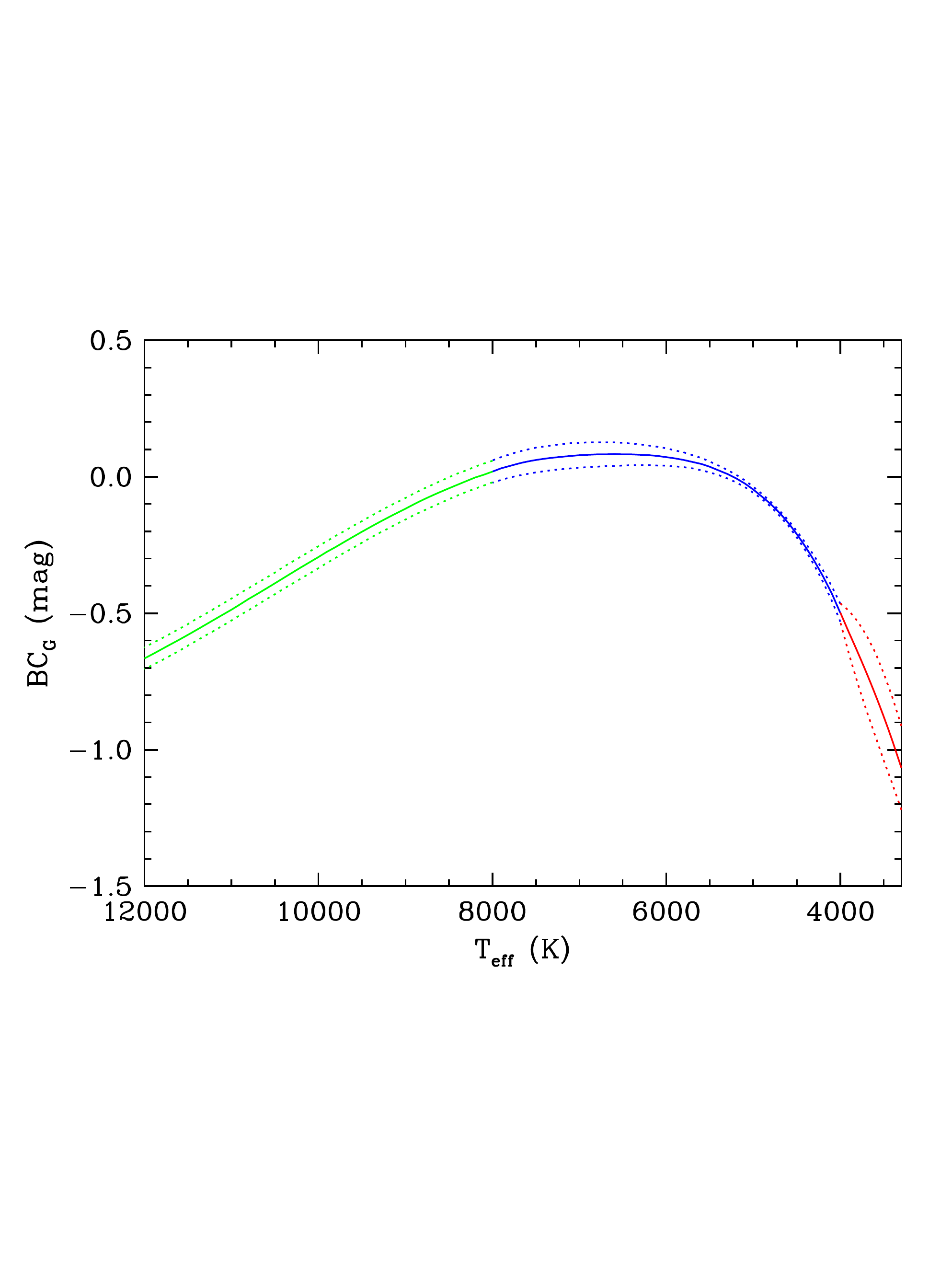}
    \caption{Bolometric corrections in the {\it Gaia\/} bandpass adopted in this work. Colors represent the three \teff\ ranges for which we have adopted our BC$_G$ versus \teff\ relations, and dotted curves represent the adopted 1$\sigma$ uncertainties for those relations.}
    \label{fig:gaia_bc}
\end{figure}

In order to develop a relation for BC$_G$ as a function of \teff\ for the widest possible \teff\ range, we have adopted the following prescription (see Figure~\ref{fig:gaia_bc}):
\begin{itemize}
\item For the range 3300--8000~K, we adopt the polynomial formulae for BC$_G$ reported by the {\it Gaia\/} team \citep[][see their Eq.~7 with coefficients in their Table~4]{Andrae:2018}, which are based on MARCS stellar atmosphere models within 0.5~dex of solar metallicity. Those relations come in two parts: 3300--4000~K\footnote{There appears to be a discrepancy between the polynomial relation for the cooler segment by \citet{Andrae:2018} and what is shown in their Figure~\ref{fig:gaia_bc}. In that figure, the BC$_G$ values extend a bit more negative at the cool end (to about $-1.7$) than indicated by their polynomial. We have not attempted to reconcile this discrepancy, but simply note it here for completeness.}, and 4000--8000~K. We have added a minor correction to the cooler segment---a shift of $+$0.0036~mag in the $a_0$ coefficient---in order to achieve continuity between the two segments. \citet{Andrae:2018} also provide additional polynomials to describe the error in BC$_G$ as a function of \teff, based essentially on the scatter as a function of \logg; we adopt these errors as well.
\item Above 8000~K, and up to the 12000~K limit available in the PHOENIX library of stellar atmosphere models, we fit a cubic polynomial to the bolometric corrections from the models, restricted to metallicities within 0.5~dex of solar, as above. We chose to consider only \logg\ values above 3.0, as our bolometric corrections here are intended for deriving radii of stars in the CTL only, from which we intentionally exclude giants. We shifted this polynomial by $+$0.01036~mag to match up exactly with the one from {\it Gaia\/} at 8000~K. For this hotter segment, we have adopted a constant error in BC$_G$ of 0.04~mag based on the scatter as a function of \logg, which also provides continuity with the {\it Gaia\/} uncertainties.
\end{itemize}

The complete set of relations described above are therefore as follows:

\noindent For the range 3300--4000~K, and where $X \equiv \teff - 5772$~K:
\begin{equation}
{\rm BC}_G = 1.7454 + 1.977\times10^{-3} X + 3.737\times10^{-7} X^2 - 8.966\times10^{-11} X^3 - 4.183\times10^{-14} X^4 \, {\rm mag} \, .
\end{equation}
The uncertainty in BC$_G$ is given by
\begin{equation}
\sigma_{{\rm BC}_G} = -2.487 - 1.876\times10^{-3} X + 2.128\times10^{-7} X^2 + 3.807\times10^{-10} X^3 + 6.570\times10^{-14} X^4 \, {\rm mag} \, .
\end{equation}

\noindent For the range 4000--8000~K:
\begin{equation}
{\rm BC}_G = 0.0600 + 6.731\times10^{-5} X - 6.647\times10^{-8} X^2 + 2.859\times10^{-11} X^3 - 7.197\times10^{-15} X^4 \, {\rm mag}
\end{equation}
with an uncertainty given by
\begin{equation}
\sigma_{{\rm BC}_G} = 2.634\times10^{-2} + 2.438\times10^{-5} X - 1.129\times10^{-9} X^2 - 6.722\times10^{-12} X^3 + 1.635\times10^{-15} X^4 \, {\rm mag} \, .
\end{equation}

\noindent For the range 8000--12000~K:
\begin{equation}
{\rm BC}_G = -3.70485 + 1.32935 Y - 0.144609 Y^2 + 0.00457793 Y^3 \, {\rm mag}
\end{equation}
where $Y \equiv \teff/1000$, with a constant uncertainty in BC$_G$ of 0.04~mag. Note that the independent variable is different here, for numerical reasons.

This last polynomial should not be extrapolated beyond 12000~K, so we are not able to compute BC$_G$ (and therefore radius using the parallax) for $\teff > 12000$~K. For stars cooler than 3300~K the {\it Gaia\/} polynomial relation could in principle be extrapolated by a small amount, however in practice we adopt the stellar parameters for M-dwarfs from the specially curated Cool Dwarf list. 

We can infer stellar mass from \teff\ for stars that are on the main sequence or not too far evolved from it. Therefore, we only apply our \teff-mass relation if the stellar radius places the star below the red giant branch, and above the white dwarf sequence, as defined in Section~\ref{subsec:assembly} (see Figure~\ref{fig:rad_vs_teff_cut}).  
Note that we implicitly are reporting a mass for stars that are subgiants; these should be regarded with caution. However, the luminosities that are reported for these subgiants are expected to be reliable, as the luminosities depend only on radius and \teff\ (see Section~\ref{sec:consistency}). 

We have revised slightly the spline relations that we developed for stellar mass as a function of \teff\ by \citet{Stassun:2018}, with the result that the formal errors are now somewhat smaller. Table~\ref{tab:mass_relation} gives the spline nodes for the mean relation \citep[unchanged from TICv7][]{Stassun:2018} and the new nodal points for the lower and upper error bars, as functions of \teff. Approximate spectral types are also provided for convenience.


\begin{center}
\vspace{-0.1in}
\begin{longtable}[c]{|c|c|c|c|c|}
 \hline
 Approx.\ Spectral Type & \teff & Mean Mass & Lower Limit & Upper Limit \\
 \hline
    .. & 55000  &  91.052 & 81.0 &  100.5\\
    O5 & 42000  &  40.0   & 36.0  &  44.0\\
    B0 & 30000  &  15.0   & 13.5  &  17.0\\ 
    .. & 22000  &   7.5   &  6.7  &   8.5\\ 
    B5 & 15200  &   4.4   &  3.95 &   4.95\\ 
    B8 & 11400  &   3.0   &  2.65 &   3.4\\ 
    A0 &  9790  &   2.5   &  2.2  &   2.85\\ 
    A5 &  8180  &   2.0   &  1.75 &   2.35\\
    F0 &  7300  &   1.65  &  1.45 &   2.00\\
    F5 &  6650  &   1.4   &  1.23 &   1.70\\
    G0 &  5940  &   1.085 &  0.965 &  1.22\\
    G5 &  5560  &   0.98  &  0.87  &  1.11\\
    K0 &  5150  &   0.87  &  0.78  &  0.98\\
    K5 &  4410  &   0.69  &  0.615 &  0.77\\
    M0 &  3840  &   0.59  &  0.51  &  0.662\\
    M2 &  3520  &   0.47  &  0.395 &  0.535\\
    M5 &  3170  &   0.26  &  0.21  &  0.30\\
    .. &  2800  &   0.117 &  0.091 &  0.14\\
    .. &  2500  &   0.056 &  0.042 &  0.07\\
 \hline
 \caption{Updated spline relation for TICv8}
 \label{tab:mass_relation}
\end{longtable}
\vspace{-0.5in}
\end{center}

\subsection{Ensuring Internal Consistency in Derived Quantities}\label{sec:consistency}
As described in the preceding sections, the basic stellar parameters that we determine for as many stars as possible are \teff\ and radius, and we also then determine mass from \teff\ where possible. To ensure that other reported stellar properties that are physically defined based on \teff, radius, and/or mass, we always calculate those dependent quantities even when empirical measures are available from other catalogs. In particular, \logg\ and mean density are always calculated from the mass and radius that we have determined. Similarly, we always calculate \lbol\ from the \teff\ and radius that we have determined.

\section{The Candidate Target List (CTL)}\label{sec:ctl}

The purpose of the CTL is to provide a subset of TIC objects that can be used to select the 
target stars for TESS 2-min cadence observations
in service of the TESS mission's primary science requirements, which are: 

\begin{enumerate}
\item To search $>$200,000 stars for planets with orbital periods less than 10~d and radii smaller than $2.5\;\rearth$.
\item To search for transiting planets with radii smaller than $2.5\;\rearth$ and with orbital periods up to 120~d among 10,000 stars in the ecliptic pole regions.
\item To determine masses for at least 50 planets with radii smaller than $4\;\rearth$. 
\end{enumerate}

Given the limited number of stars for which TESS will be able to acquire 2-min cadence light curves, it is crucial that the set of targets for TESS be optimized for detection of small planets. To that end, we have compiled a catalog of bright stars that are likely to be dwarfs across the sky, from which a final target list for TESS can be drawn, based on in-flight observation constraints. 
This list of high-priority candidate 2-min cadence targets is the CTL. Our basic consideration is to assemble a list of dwarf stars all over the sky in the temperature range of interest to TESS, bright enough for TESS to observe, and taking extra steps to include the scientifically valuable M dwarfs. 

Our overall approach is to start with the $\sim$1.7 billion stars in the TIC, and then apply cuts to select stars of the desired ranges in apparent magnitude and spectral type, and to eliminate evolved stars. At this stage we also compute additional information that is relevant for target selection, which for logistical reasons or computational limitations, we do not compute for all other stars in the TIC.

First, we give a brief overview describing the assembly of the CTL from the TIC, including specifically the process by which we identify likely dwarf stars for inclusion in the CTL and identify likely red giants and white dwarfs for exclusion from the CTL.
Next we describe the algorithms by which we calculate improved measures of uncertainties on the stellar parameters and flux contamination in the expected photometric aperture of each star (Section~\ref{sec:ctl_algorithms}). Finally, we present the prioritization scheme used for identifying the top priority targets from the CTL for targeting (Section \ref{subsec:priority}). 
The CTL is provided for use through MAST and for interactive use via the Filtergraph data visualization system \citep{Burger:2013} at {\tt \url{http://filtergraph.vanderbilt.edu/tess_ctl}}.
A summary of the quantities included in the CTL on the Filtergraph portal is provided in Appendix~\ref{sec:appendix_filtergraph}. 

\subsection{Selection of Target Stars for the CTL\label{subsec:assembly}}

From the $\sim$1.7 billion point sources in the TIC we initially select stars for the CTL if they: (1) have parallaxes and $G G_{RP} G_{BP}$ photometry reported by {\it Gaia\/} DR2 that satisfy quality criteria on reduced $\chi^2$, number of degrees of freedom, photometric excess factor, and on the $G$ and $G_{BP} - G_{RP}$ colors \citep[see equations 1 and 2 in][]{Arenou:2018}; 
and (2) satisfy the condition $T<13$. We implement the \tmag\ criterion to reduce the CTL to a manageable size, emphasizing the bright dwarfs that are likely to be the highest priority targets.
Note that while this \tmag\ cut would by itself eliminate many M-dwarfs, we rely on the specially curated Cool Dwarf List to ensure the inclusion of high-priority, bona fide M-dwarfs.

\begin{figure}[!ht]
    \centering
    \includegraphics[width=0.765\linewidth]{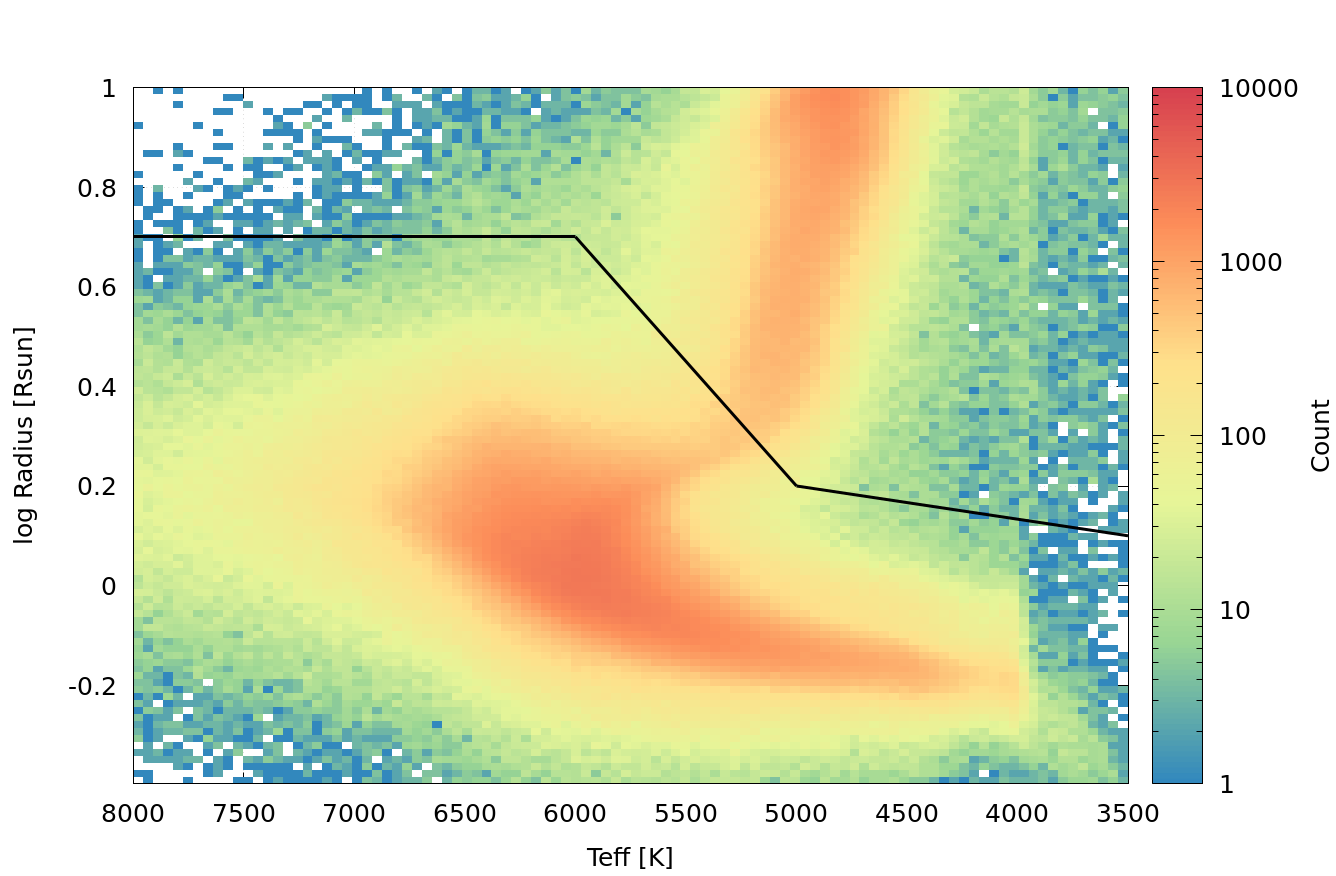}
    \includegraphics[width=0.5\linewidth]{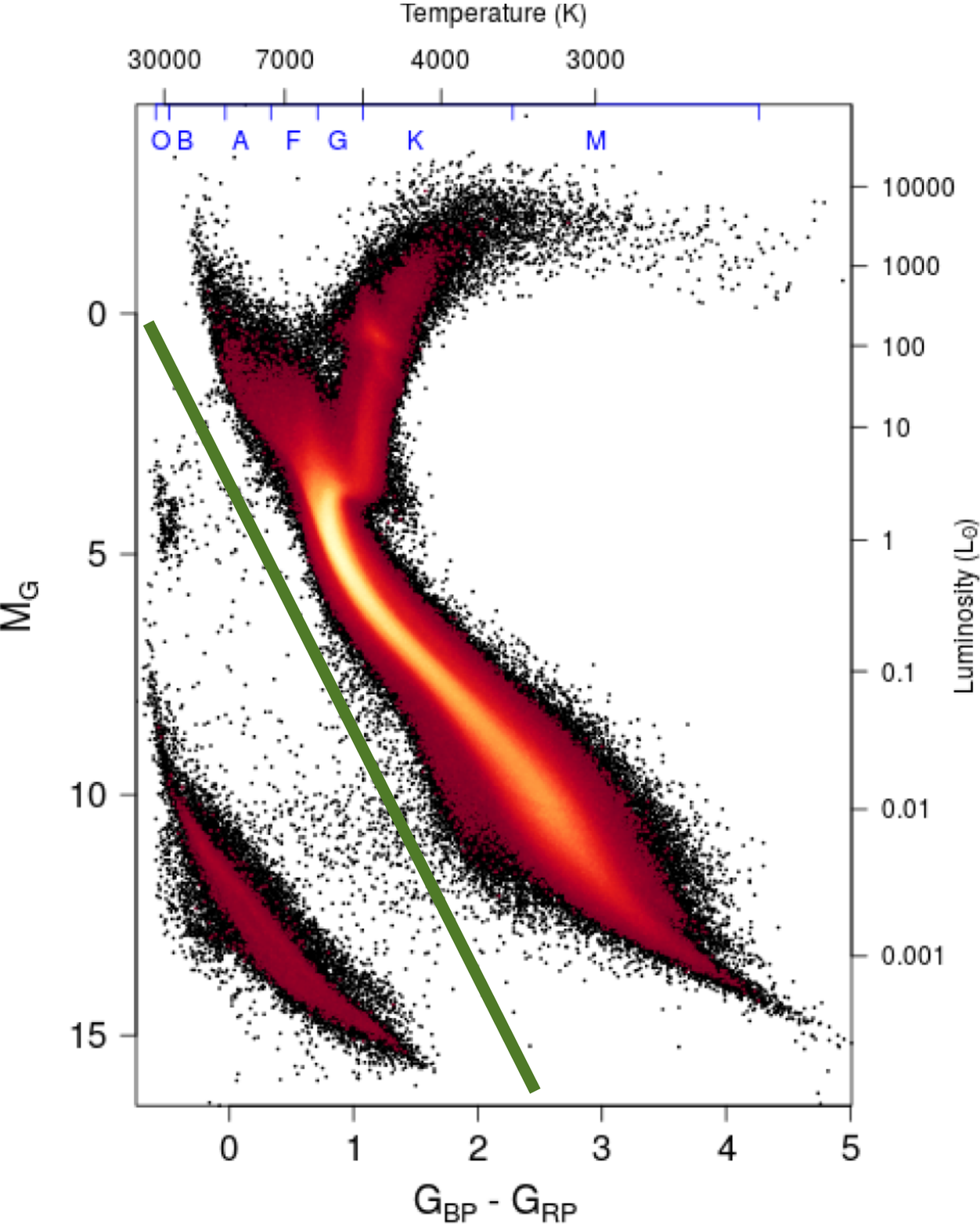}
    \caption{\textit{Top}: Radius versus \teff\ for stars with calculated parameters in the TIC, showing the basis for the radius cuts adopted to include dwarfs and subgiants but exclude red giants from the calculation of mass and \logg. \textit{Bottom}: $M_{G}$ vs.\ $\gbp-\grp$ diagram reproduced from Figure~2 of \citet{Gaia:2018v}, with a line drawn by eye showing the basis for the color-magnitude cut adopted to exclude white dwarfs from the calculation of mass and \logg. The equation for this line is $M_G = 5.15 (\gbp-\grp) + 4.12$.}
    \label{fig:rad_vs_teff_cut}
\end{figure}

Next, 
we cut on stellar radius to eliminate red giants, as shown in Figure~\ref{fig:rad_vs_teff_cut}.
Note that this explicitly includes subgiants ($3.5 < \logg < 4.1$); recognizing that some subgiants can be considered high-value in some cases, we include them but rely on the target prioritization metric and its dependence on stellar radius (see below) to ensure that bright subgiants do not overwhelm the selection of final 2-min cadence targets. The specific radius cuts adopted as a function of \teff\ are as follows (see Figure~\ref{fig:rad_vs_teff_cut}): 
For $\teff \ge 6000$~K, the dividing line is $\log R/R_\odot = 0.7$, then the nodes for the subsequent piecewise linear dividing lines are (\teff, $\log R/R_\odot$) = (5000~K, 0.2) and (2000~K, 0.0). 

To exclude stars likely to be white dwarfs we used a diagram of absolute $G$ magnitude versus $\gbp-\grp$ color (Figure~\ref{fig:rad_vs_teff_cut}, bottom panel), which shows the main sequence and white dwarf sequences clearly separated. We defined a boundary by eye, represented by the equation $M_G = 5.15 (\gbp-\grp) + 4.12$ and shown by the line in Figure~\ref{fig:rad_vs_teff_cut}. We eliminated stars below this boundary (after proper corrections for reddening) as being probable white dwarfs.

\begin{figure}[!ht]
    \centering
    \includegraphics[width=\linewidth]{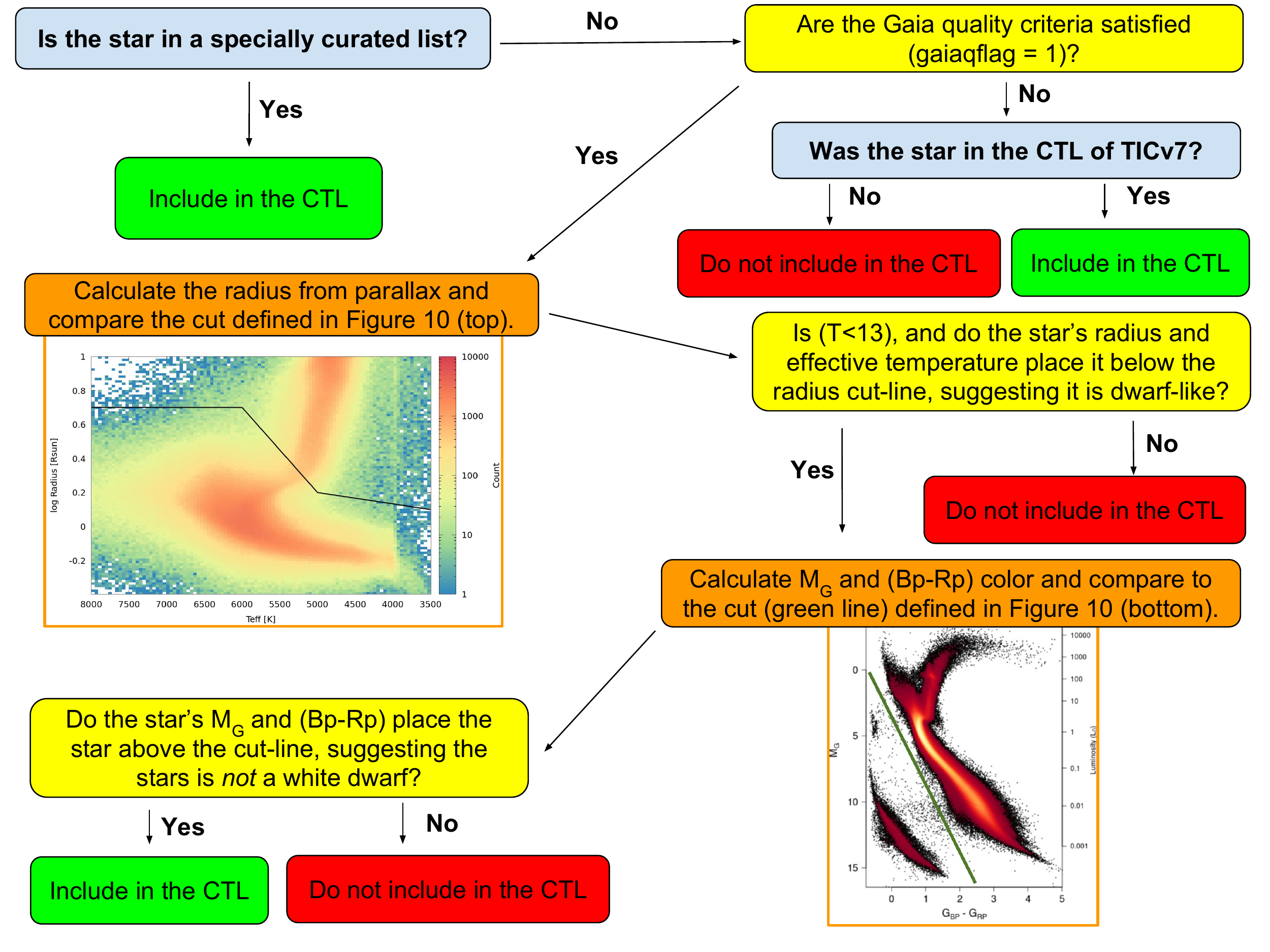}
    \caption{A visual schematic of the logic flow by which stars are selected from the TIC into the CTL.}
    \label{fig:giant_removal}
\end{figure}

The entire procedure described above is summarized in logical flowchart form in Figure~\ref{fig:giant_removal}. 
We do not include stars in the CTL if we are unable to determine their \teff\ spectroscopically or from dereddened colors (see Sec.~\ref{sec:dereddening}), or if we are unable to estimate their radius (Sec.~\ref{sec:mass_radius}) or the flux contamination from nearby stars (Sec.~\ref{sec:flux_contam}) since these are essential to setting target priorities (see Sec.~\ref{subsec:priority}). 
All stars in the specially curated Cool Dwarf and Hot Subdwarf target lists (Appendix~\ref{sec:lists}) are included in the CTL.  
Finally, in order to ensure inclusion of high-priority stars that may be missing from {\it Gaia} DR2, stars previously included in the CTL of TICv7 
on the basis of a reduced proper-motion cut suggesting that they are dwarfs, and
for which {\it Gaia\/} DR2 does not provide sufficiently reliable information to warrant their exclusion (according to the quality criteria discussed above), are included in the CTL.
The CTL at present comprises 9.48~million stars. 

Strictly speaking, the CTL as delivered to NASA is simply a list of candidate target stars with associated relative targeting priorities. We are providing an enhanced version of the CTL with all relevant observed and derived stellar quantities described here, through the Filtergraph Portal system as a tool for the community to interact with this unique data set. Appendix~\ref{sec:appendix_filtergraph} describes each quantity in the CTL that can be found on the Filtergraph Portal system.

\subsection{Algorithms for calculated stellar parameters \label{sec:ctl_algorithms}}

\subsubsection{Flux contamination\label{sec:flux_contam}}
We follow the same procedures as in the original CTL \citep{Stassun:2018}, with the same assumed parameters. Briefly, contaminants are searched for within 10 TESS pixels of the target, and the contaminating flux is calculated within a radius that depends on the target's TESS magnitude, and uses a PSF that is based on pre-launch PSF measurements of the field center from the SPOC (note that the PSF model does not attempt to account for bleed trails from very bring stars). The flux contamination reported is simply the ratio of the total contaminant flux to the target star flux. See Section~3.2.3 of \citet{Stassun:2018} for more details.

\begin{figure}[!ht]
    \centering
    \includegraphics[width=0.8\linewidth,trim=0 0 0 70,clip]{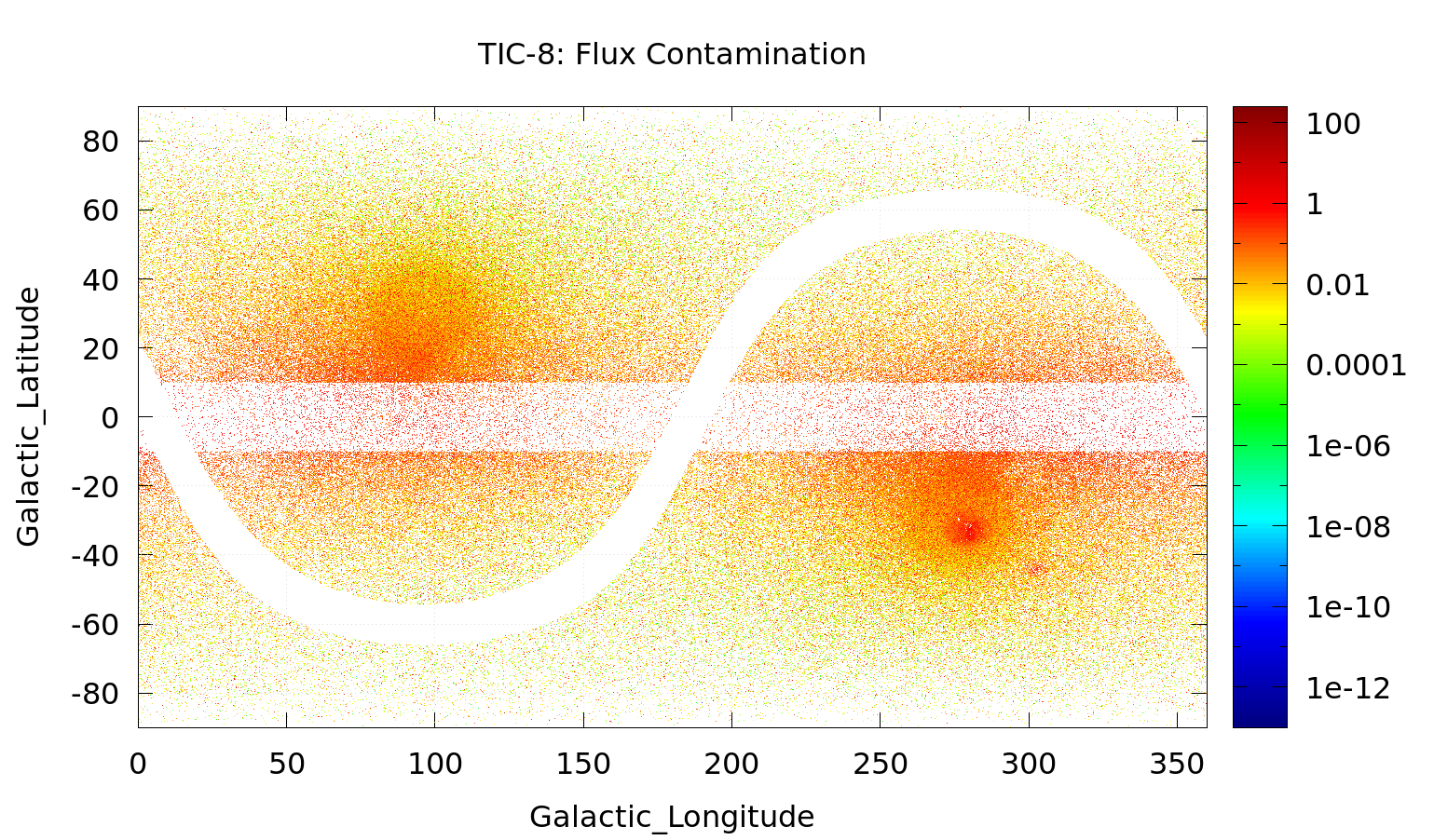}
    \caption{Flux contamination as a function of sky position in TICv8.}
    \label{fig:flux_contam}
\end{figure}

\subsubsection{Monte Carlo Determination of Parameter Uncertainties}

We have implemented a Monte Carlo based approach to improve the final uncertainty estimates for the stellar parameters reported in the CTL, in which we perturb each observed quantity 1000 times and carry the perturbed values through the calculations to obtain a distribution for each derived quantity. We then report the 16th and 84th percentiles of those distributions as the corresponding lower and upper error bars. We have done this for two main reasons. First, we wish to be able to report asymmetric errors to better reflect the nature of the parameter posteriors. Second, a simple summation in quadrature of the underlying parameter errors overestimates the final uncertainty. This overestimation of the final uncertainty is particularly severe for the stellar radius because: 
The distance error enters four times ($D$, $G$, $T_{\rm eff}$, BC$_G$); 
the errors from the reddening maps enter three times ($G$, $T_{\rm eff}$, BC$_G$); 
the error from the \teff\ calibration enters twice ($T_{\rm eff}$, BC$_G$);
and 
photometric errors in $G_{BP} - G_{RP}$ enter twice ($T_{\rm eff}$, BC$_G$).


In what follows, we represent Monte Carlo perturbed quantities with primed (or double-primed) symbols and nominal values with unprimed symbols. In addition, $\mathcal{N}$ represents a normal Gaussian deviate (mean = 0, $\sigma = 1$) that is used to perturb the nominal quantities. Once a quantity is perturbed, we use the same perturbed value throughout the procedure in order to preserve parameter correlations. For perturbing quantities with asymmetric error bars we assume each side is reasonably well represented by a Gaussian distribution, and use the lower error bar if the Gaussian deviate is negative or the upper error bar otherwise.

The procedure follows the steps below, in sequence: 
\begin{enumerate}
\item  Perturb the stellar distance: $D' = D + \mathcal{N} \times \{\sigma_{D,{\rm low}}, \sigma_{D,{\rm high}} \}$
\item  Perturb the reddening, which involves two contributions: one from the distance, and another from the intrinsic dust map errors. Query the dust map(s) with $D'$ to obtain a perturbed reddening $E(B-V)''$. For Pan-STARRS, find the 16th and 84th percentiles of the reddening distribution at the nominal distance, and subtract from reddening at nominal distance to obtain the intrinsic dust map errors, \{$\sigma_{\rm red,low}$, $\sigma_{\rm red,high}$\}. For Schlegel, adopt $\sigma_{\rm red,low} = \sigma_{\rm red,high} = 0.01$ mag. Then compute $E(B-V)' = E(B-V)'' + \mathcal{N}\times \{\sigma_{red,low}, \sigma_{red,high} \}$, and calculate reddening and extinction in the {\it Gaia\/} passbands with $E(\gbp-\grp)' = 1.31 E(B-V)'$ and $A(G)' = 2.72 E(B-V)'$.
\item  Perturb the {\it Gaia\/} color and magnitude using the photometric errors:
$(\gbp-\grp)' = (\gbp-\grp) + \mathcal{N} \times \sigma_{\gbp} + \mathcal{N}\times \sigma_{\grp}$, and $G' = G + \mathcal{N} \times \sigma_G$. 
\item  Deredden the perturbed {\it Gaia\/} color and magnitude:
$(\gbp-\grp)'_{\rm dered} = (\gbp-\grp)' - E(\gbp-\grp)'$ and $G'_{\rm dered} = G' - A(G)'$. 
\item  Compute the perturbed and dereddened TESS magnitude $T'_{\rm dered}$ with the expression in Section~\ref{subsubsec:tmag} using $G'_{\rm dered}$ and $(\gbp-\grp)'_{\rm dered}$. Calculate the perturbed extinction in the $T$ band as $A(T)' = 2.06 E(B-V)'$. Then apply the extinction and compute the final perturbed apparent TESS magnitude (i.e., affected by extinction) as $T' = T'_{\rm dered} + A(T)' + \mathcal{N}\times \sigma_T$,
where $\sigma_T$ is the scatter of the $T$ calibration. 
\item  Compute the perturbed \teff: $\teff' = T_{\rm eff,dered}' + \mathcal{N} \times \sigma_{T_{\rm eff}}$,  where $T_{\rm eff,dered}'$ is the perturbed temperature derived from the perturbed dereddened color, and $\sigma_{T_{\rm eff}}$ = 122~K (Section~\ref{subsubsec:teff}). 
\item  Compute the perturbed radius ($R'$) from the perturbed bolometric correction (${\rm BC}'$), $\teff'$, $D'$, and $G'_{\rm dered}$, where ${\rm BC}' = {\rm BC}(\teff') + \mathcal{N}\times \sigma_{\rm BC}$, and $\sigma_{\rm BC}$ is a function of $\teff'$. 
\item  Compute the perturbed mass as
$M' = M(\teff') + \mathcal{N}\times \{\sigma_{M,{\rm low}}, \sigma_{M,{\rm high}} \}$, where $\sigma_M$ is a function of $\teff'$. Then compute the perturbed \logg, luminosity, and mean density as $\logg' = 4.4383 + \log M' - 2\log R'$, $L' = (R')^2 (\teff'/5772)^4$, and $\rho' = M' / (R')^3$.
\end{enumerate}

\subsection{Target prioritization\label{subsec:priority}}
Ultimately, one of the most fundamental characteristics reported in the CTL is the target priority, which allows the selection of the most suitable stellar candidates for 2-minute cadence. The target priority calculated in the CTL of TICv8 uses an identical schema to the priority calculation in the CTL of TICv7. For a more detailed derivation of the priority formula, we direct the reader to \citet{Stassun:2018}. However, we provide a basic explanation below.

The priority in TICv8 determines the relative ability of TESS to detect small planetary transits, and is calculated using the radius of the star ($R$), and the total expected photometric precision ($\sigma$), and a priority boost factor which scales with a probabilistic model of the expected number of sectors ($N_S$) any given star could fall in. Typically, the closer the star is to the Ecliptic North or South pole, the larger the boost factor. This leads to the following formulation of stellar priority: 
\begin{equation}
    \frac{\sqrt{N_S}}{R^{1.5}\times\sigma}
\end{equation}

This priority is then normalized by the priority for a star with $R = 0.1 R_{\odot}$, $N_S = 12.654$ sectors, and $\sigma = 61.75$~ppm to force the priority to be on a scale from 0 to 1. Finally, there are a small subset of stars that are manually de-prioritized based on known issues with current TIC calculations, or known limitations of the \textit{TESS} observing plan. These are:

\begin{itemize}
    \item Stars close to the Galactic Plane ($|b| <10^{\circ}$) are multiplied by a factor of 0.1 in order to de-prioritize stars that may be affected by a poor understanding of their true reddening. Stars in the specially curated lists are excluded from this condition. 
    \item Stars with \logg~values greater than 5 have had their priorities set to 0 and their properties set to Null, to avoid biases from poor quality effective temperature, extinction, or parallax measurements. Stars in the specially curated lists are excluded from this condition.
    \item Stars close to the Ecliptic Plane ($|\beta | \lesssim 6^{\circ}$) are not expected to be observed as part of the main mission due to a gap in camera coverage between the Southern and Northern observations. Therefore, their $N_S$ values are 0, and thus the priority is 0.
\end{itemize}



\section{Summary of Representative Properties of Stars in the TIC}\label{sec:discussion}

Compared to TICv7, the number of stars in TICv8 has increased by a factor of $\sim$3.5. The number of stars with \teff\ has doubled, and the number with estimated radii has increased by a factor of $\sim$20. Table~\ref{tbl:summary} summarizes the numbers of stars in the TICv8 and CTLv8 for various representative subsets. 

\begin{figure}[!ht]
    \centering
    \includegraphics[width=0.9\linewidth]{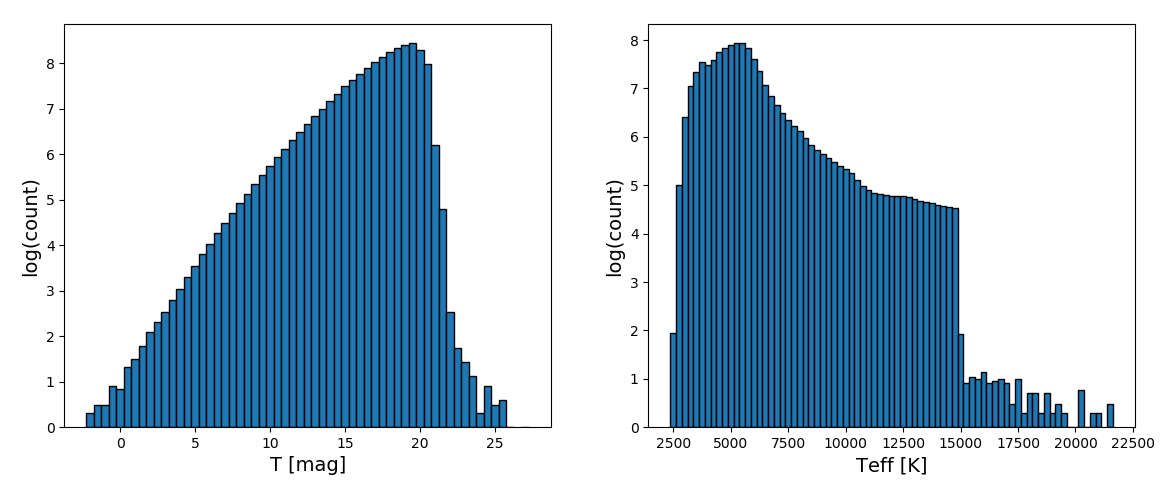}
    \caption{TICv8 distributions of \tmag\ (left) and \teff\ (right).}
    \label{fig:tmagteffhist}
\end{figure}

\begin{figure}[!ht]
    \centering
    \includegraphics[width=0.9\linewidth]{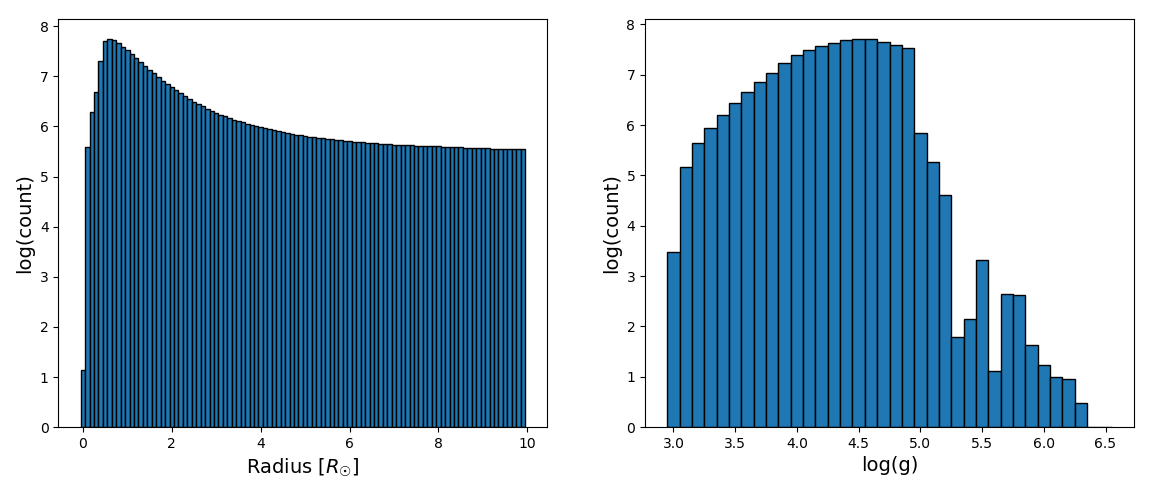}
    \caption{TICv8 distributions of stellar radius for stars smaller than 10~\rsun\ (left) and \logg\ (right).}
    \label{fig:radlogghist}
\end{figure}

Figure~\ref{fig:tmagteffhist} shows the overall distribution of TESS magnitudes (left) and \teff\ (right). Note that our relations for estimating \teff\ end at $15000$~K; hotter stars in the distribution originate from the specially curated list for hot subdwarfs. 
Figure~\ref{fig:radlogghist} shows the overall distribution of radii for stars smaller than 10~\rsun\ (left) and $\log g$ (right). Stars with $R \ge 5$~\rsun\ are regarded as giants and we do not report masses or other derived properties for them. Our relations for masses are designed principally for dwarfs, and work reliably well also for subgiants, but are not reliable for giants. 

\begin{center}
\vspace{-0.1in}
\begin{longtable}[c]{|c|c|c|c|c|c|c|}
 \hline
 Quantity & \multicolumn{2}{c|}{Number of Stars} & Sub-population & \multicolumn{3}{c|}{Number of Stars} \\
    & TICv7 & TICv8 & & TICv7 & TICv8 & CTLv8  \\
 \hline
 \tmag\ magnitude & 470,995,593 & 1,726,340,024 & $\tmag < 10$ & 966,297 & 912,552 & 268,752\\
 \teff & 331,414,942 & 683,248,319 &  $\teff < 4500$~K & 991,868 & 140,614,051 & 4,053,071\\
 Radius & 27,302,067  & 541,007,000 & $R < 0.5$\;\rsun & 787,924 & 27,804,756 & 1,568,574\\
 Mass & 27,302,066 & 455,211,680  &  $M < 0.5$\;\msun & 741,483 & 14,113,970 & 1,587,663\\
 Spectroscopic \teff\  & 572,363  & 4,059,381 & Spect. \teff\ $<6000$ and $\logg > 4.1$ & 395,144 & 1,673,350 & 420,443\\ 
 Proper motion & 316,583,013 & 1,335,789,302 & Proper Motion $>1000$~mas~yr$^{-1}$ &  655 & 1092 & 498\\
 Parallax & 2,045,947 & 1,269,096,797 & Distance $<100$~pc & 42,454 & 574,927 & 217,245\\
 \hline
 \caption{Summary of basic stellar properties in the TIC and CTL. Note that the TICv7 sub-populations for cool \teff, small radii, and low mass reflect numbers from CTLv7, because in TICv7 these quantities were computed only for stars in the CTL. 
 \label{tbl:summary}}
\end{longtable}
\vspace{-0.45in}
\end{center}

\acknowledgments
We gratefully acknowledge partial support for this effort from NASA grant 17-XRP17 2-0024. 
Funding for the TESS mission is provided by NASA's Science Mission directorate.
P.S.M.\ acknowledges support from the NASA Exoplanet Research Program (XRP) under Grant No.~NNX15AG08G issued through the Science Mission Directorate.
B.R-A.\ acknowledges funding support from CONICYT PAI/CONCURSO NACIONAL INSERCIÓN EN LA ACADEMIA, CONVOCATORIA 2015 79150050 and FONDECYT through grant 11181295.
This work is partly supported by JSPS KAKENHI Grant Numbers JP18H01265 and JP18H05439, and JST PRESTO Grant Number JPMJPR1775.
This work has made extensive use of the Filtergraph data visualization service \citep{Burger:2013}.
This work has made use of data from the European Space Agency (ESA) mission {\it Gaia} (\url{https://www.cosmos.esa.int/gaia}), processed by the {\it Gaia} Data Processing and Analysis Consortium (DPAC, \url{https://www.cosmos.esa.int/web/gaia/dpac/consortium}). Funding for the DPAC has been provided by national institutions, in particular the institutions participating in the {\it Gaia} Multilateral Agreement. This work has also made use of the SIMBAD database and the VizieR catalog access tool, both operated at the CDS, Strasbourg, France, and of NASA's Astrophysics Data System Abstract Service.

\clearpage
\appendix
\section{Specially Curated Lists}\label{sec:lists}

The specially curated lists from TICv7 \citep{Stassun:2018} have been updated as follows. 

\subsection{Cool Dwarf List}

The Cool Dwarf List has been updated. It is incorporated into the TIC and CTL as a total override, meaning that values in this list supersede and replace default values calculated by the usual TIC/CTL procedures. Muirhead et al.\ (in preparation) provides detailed procedures. Here we briefly summarize the main changes compared to the Cool Dwarf List that was incorporated into the previous version of the TIC/CTL \citep{Stassun:2018,Muirhead:2018}.

For TICv8, the Cool Dwarf specially curated list was revised and substantially augmented to include newly available astrometric parallax measurements and photometry from the {\it Gaia\/} Mission.  The Second Cool Dwarf Catalog (CDC2) was built from the ``nearest neighbor" cross-match between {\it Gaia\/} DR2 sources and the Two Micron All Sky Survey Point Source Catalog \citep[2MASS PSC;][]{Cutri:2013}, available on the {\it Gaia\/} Archive.  The cross-matched catalog was queried for all objects with the following criteria:
\begin{itemize}
\item Non-zero astrometric parallax measurement with a signal-to-noise of at least 5.
\item A single and unique entry in the 2MASS PSC, and a photometric quality flag of ``C" or better for all 2MASS magnitudes.
\item Absolute $K_S$-band magnitude ($M_K$) between 4.5 and 10.0.
\item $V-J$ color greater than 2.7, to identify cool stars and maintain consistency with CDC1.
\item Absolute $V$-band magnitude ($M_V$) that meets the following criterion: $M_V > 2.2 ( V-J ) - 2$.
\item {\it Gaia\/} \grp-band magnitude less than 18.
\end{itemize}

For the absolute magnitude calculations, distances were taken from \citet{Bailer-Jones:2018}.  The $V$-band magnitude was calculated using \gbp\ and \grp\ magnitudes and the conversion published by \citet{Jao:2018}.  No extinction or reddening corrections were applied in the query.

The result from the query was cross-matched with the original Cool Dwarf Catalog \citep{Muirhead:2018}, including all entries from both catalogs.  For each entry, we calculated stellar mass, stellar radius, effective temperature and TESS magnitude, assuming each entry is a single star and ignoring effects from reddening and extinction.

Stellar masses and radii were calculated using the mass-$M_K$ relations from \citet{Mann:2019} and the radius-$M_K$ relations from \citet{Mann:2015}, both valid for $4.5<M_K<10.0$.  Objects outside of this range or lacking an astrometric parallax were flagged for removal.   Effective temperature and \tmag\ were calculated from \gbp\ and \grp\ magnitudes using custom relations developed from photometrically-calibrated spectra from \citet{Mann:2013}.  Objects in the CDC1 without unique 2MASS identifiers were flagged for removal.

\begin{figure}[!ht]
    \centering
    \includegraphics[width=0.49\linewidth]{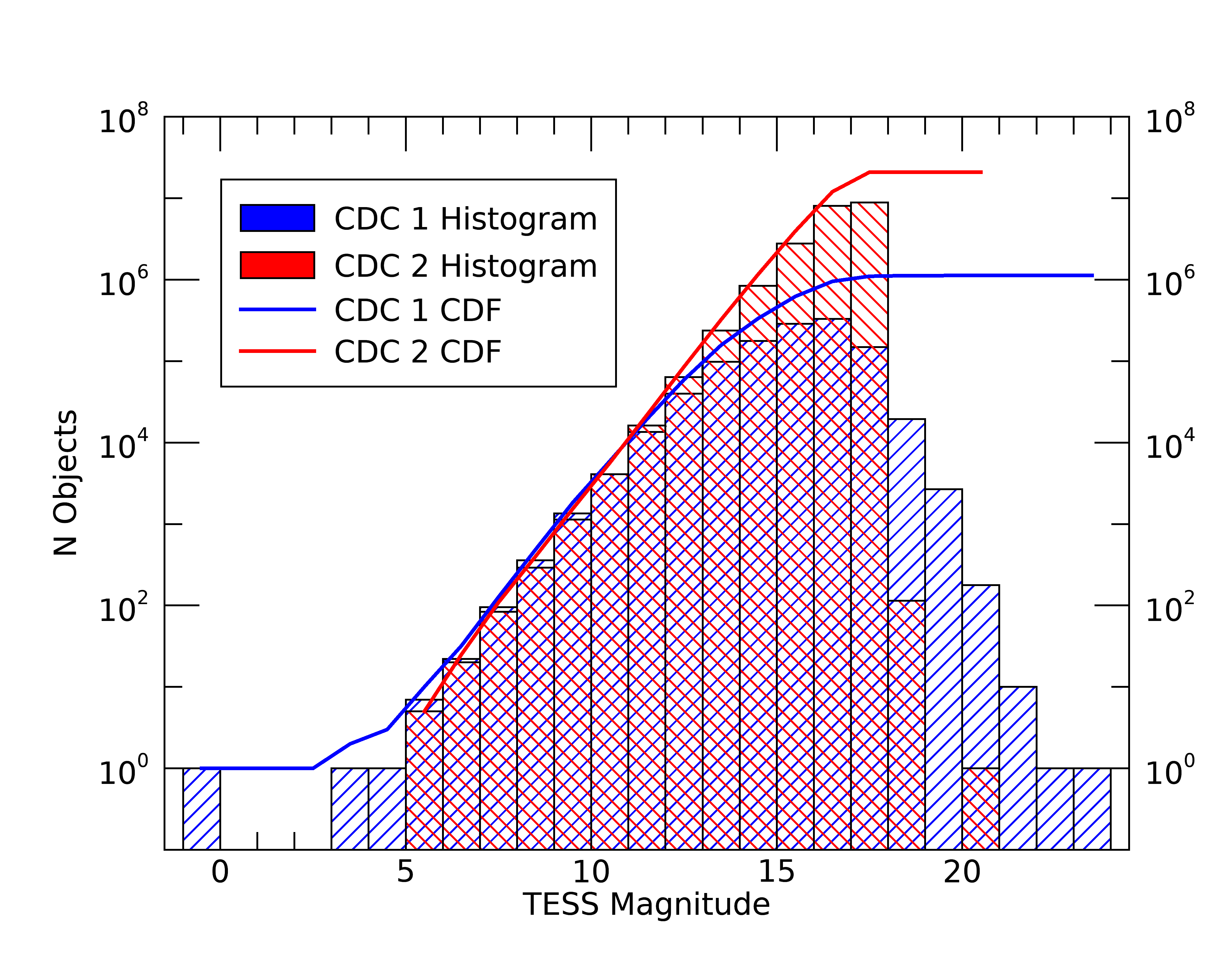}
    \includegraphics[width=0.49\linewidth]{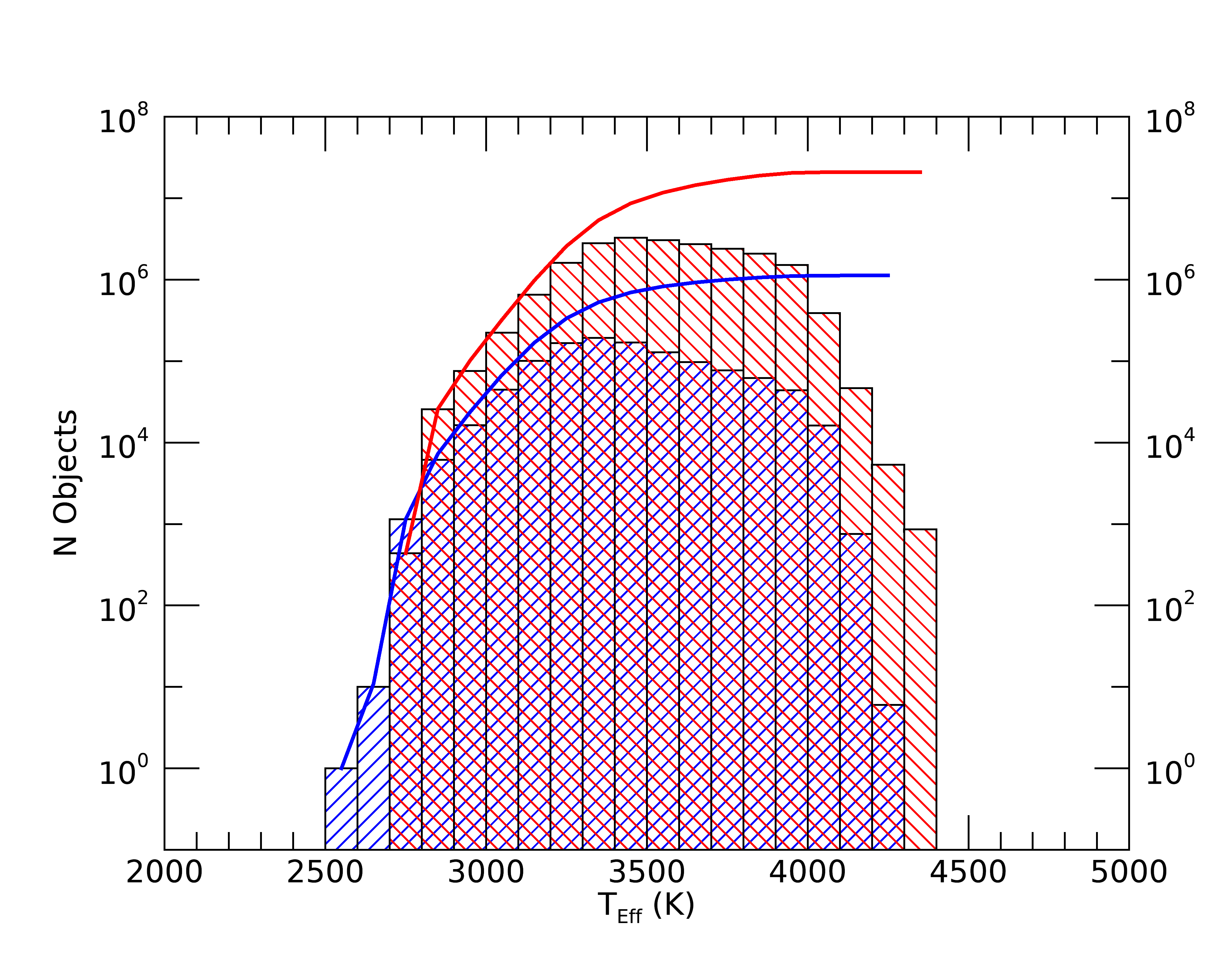}
    \caption{Histograms and cumulative distribution functions comparing the updated Cool Dwarf Catalog (CDC2) to that which was incorporated into the previous TIC/CTL (CDC1), for \tmag\ (left) and \teff\ (right). Several bright objects in CDC1 were excluded from CDC2 owing to the requirements on $M_K$. Additionally, the requirement that $\grp < 18$ reduces the number of faint cool dwarfs with \tmag\ greater than 18. CDC2 lacks objects with \teff\ less than 2700~K.}
    \label{fig:cdc_figs}
\end{figure}

\begin{figure}[!ht]
    \centering
    \includegraphics[width=0.47\linewidth]{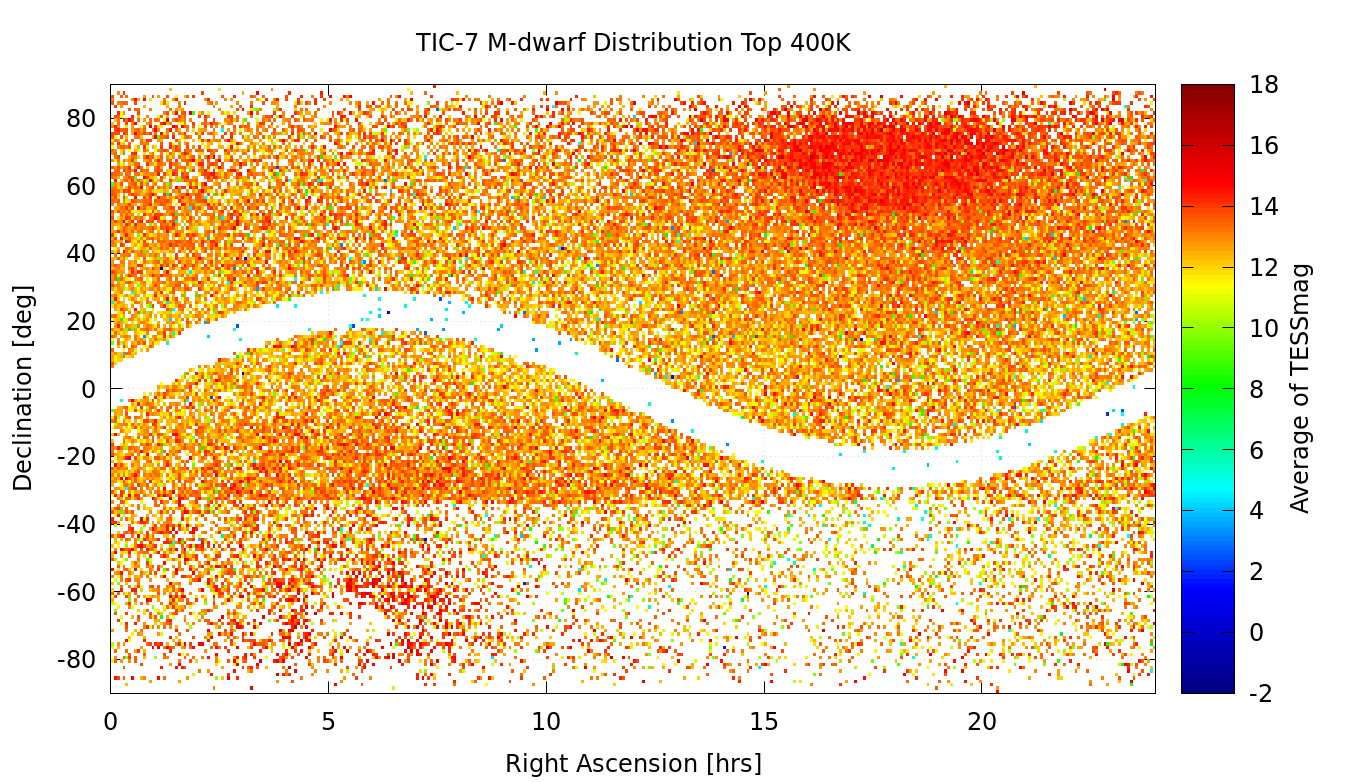}
    \includegraphics[width=0.5\linewidth]{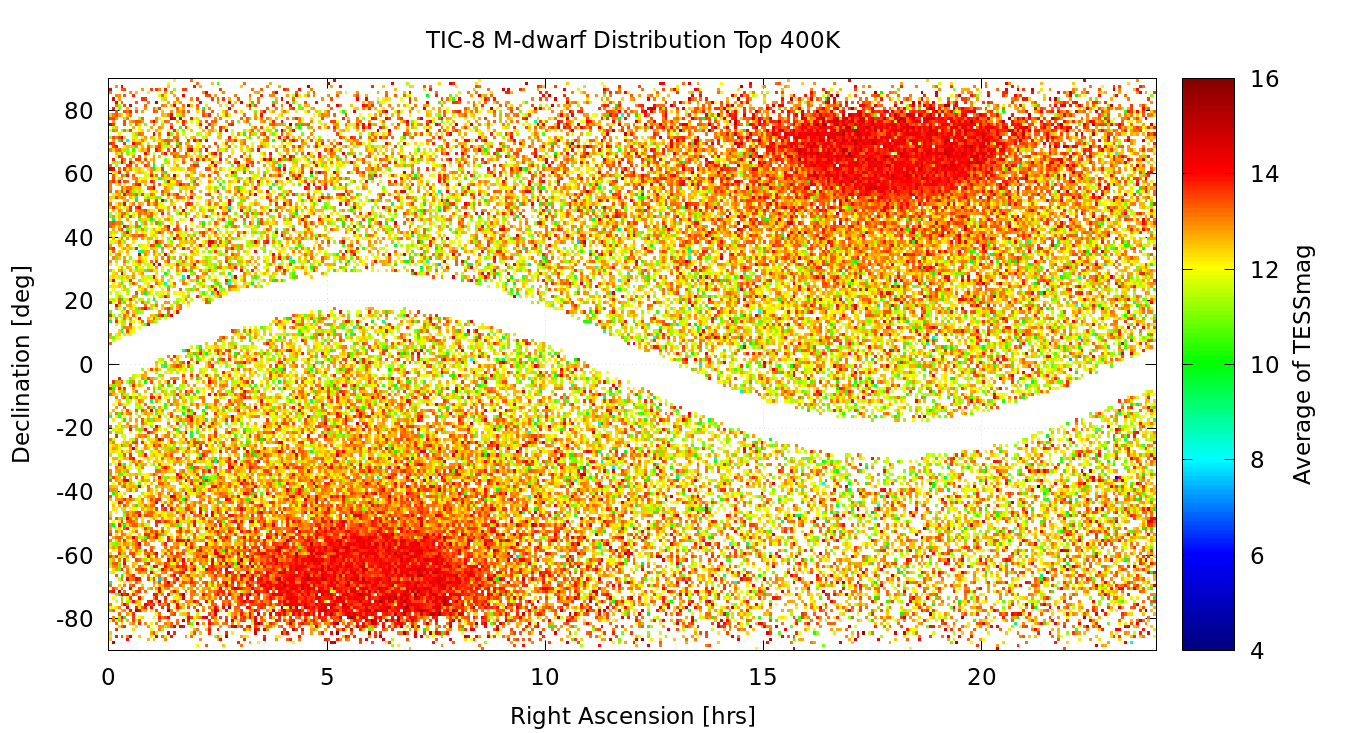}
    \caption{Comparison of the distribution of idnetified cool dwarfs in TICv7 (\textit{left}) and TICv8 (\textit{right}) in the top 400K targets.}
    \label{fig:cdc_comps}
\end{figure}

Figure~\ref{fig:cdc_figs} shows histograms and cumulative distribution functions comparing CDC1 and CDC2, for \tmag\ and \teff.  CDC2 removed a handful of bright and low-temperature CDC1 entries owing to the parallax and $M_K$ criteria.  For example, the brightest object in the CDC1 is $\alpha$~Centauri~B, a K1 dwarf that does not meet the new $M_K$ criteria. 
For context, Figure~\ref{fig:cdc_comps} compares the spatial distribution of CDC stars in CDC2 versus CDC1, showing especially the substantially improved coverage of the southern sky (including especially the southern CVZ) in CDC2. 

Due to the CDC's use of specialized relations for determining \tmag, we observe an offset of $\sim$0.1~mag between \tmag\ as computed in the CDC versus \tmag\ computed by our nominal relations (see Section~\ref{subsubsec:tmag}). 
The difference can be as large as $\sim$1~mag at the faintest end of the TIC (\tmag $\gtrsim$18). 
Because we adopt the CDC values as an override, these offsets will only be noticeable when comparing similar stars where one is in the CDC and the other is not. 

Finally, the \teff\ values computed in the CDC versus the standard TIC relations are in good agreement, especially for TIC \teff\ derived from {\it Gaia\/} colors; the scatter is even lower than the 122~K that we assume for the standard TIC \teff. However, stars with \teff\ inherited from TICv7 can differ more substantially, especially stars with high \teff\ derived using a photometric $B$ magnitude (provenance flag ``bphotvk"); these should be checked independently before being used.


\subsection{Known Planet Hosts}
There are $\sim$3000 stars known to host exoplanets, of which $\sim$2800 are systems for which the radial velocity and transit methods were used for discovery. There are a variety of scientific reasons why TESS observations of these stars would be valuable, such as detecting stellar variability \citep{Dragomir:2012}, transit ephemeris refinement for follow-up observations \citep{Kane:2009}, detecting TTVs to help identify additional planets or stellar companions, 
potential discovery of further transiting planets in those systems, 
etc.  We worked to include all known planet hosts in previous versions of the TIC, and these stars were also selected in a Cycle~1 TESS Guest Investigator program, ensuring that they are observed at 2-minute cadence.

While we have continued to make sure that all known planet hosts are included in the updated TIC, the stellar parameters of those stars were determined according to the standard procedures outlined in this paper.  Those procedures, as noted, are based on large catalogs, and do not take advantage of the precise measurement of individual systems that are typically conducted when the planets are discovered.  We explored the option to adopt a curated set of the stellar parameters of the planet hosts to improve those quantities in the TIC by incorporating the exoplanet host star parameters listed in the NASA Exoplanet Archive \citep{Akeson:2013}.  Unfortunately, that catalog, along with all other catalogs of exoplanet host parameters that we explored, are by nature somewhat incomplete and heterogeneous.  We have been unable as yet to adopt the set of host star parameters from such a catalog without requiring star-by-star customization of the stellar property fields to maintain the levels of internal consistency that the TIC itself adheres to.  We therefore decided not to adopt this information for the current TIC.

We do not expect this change in the treatment of known exoplanet hosts to have major effects on the TIC.  All such stars are still listed in the TIC, and we have no reason to believe that their stellar parameters---as determined through the procedures described in this paper along with all other TIC stars---are any less reliable than the rest of the TIC stars of similar stellar types.  Known planet host stars can still be observed for various science goals by TESS, and the only impact that the absence of highly curated parameters will present is a less precise determination of resulting transit properties from the default TESS transit search pipeline.  However, individual investigators can always recalculate such properties themselves using published stellar information.


\subsection{Other Lists}
\begin{itemize}
\item {\it Bright Stars:} No longer exists; bright stars are included in the TIC but not as a specially curated list with separate procedures nor with special priorities. 
\item {\it Hot Subdwarfs:} Has been updated; these are incorporated into the TIC and CTL as a total override.
\item {\it Guest Investigator Targets:} We import proposed GI Cycle 2 targets that did not have a pre-existing TIC-ID as new objects. 
\end{itemize}

\section{Provenance Flags in the TIC\label{sec:flags}}
\subsection{Provenance Flags in earlier versions of the TIC}
\begin{center}
\begin{longtable}[c]{|rlll|}
\hline 
Column & Name & Flags & Description  \\
\hline 
12            & Objtype & ... &  Flag to identify the object's type \\
    ...       &  ...   & star  & object is a star \\
    ...       &  ...   & extended  & object is a galaxy/extended source \\
    
13            & Typesrc & ...  & Flag to identify the source of the object\\  
    ...       &  ...   & gaia2  & stellar source from Gaia DR2 \\
    ...       &  ...   & hip  & stellar source is hipparcos \\
    ...       &  ...   & cooldwarfs  & stellar source is the cool dwarf list \\
    ...       &  ...   & 2mass  & stellar source is 2MASS \\
    ...       &  ...   & lepine  & stellar source is Lepines All-sky Catalog of Bright M Dwarfs (2011) \\
    ...       &  ...   & tmgaia & stellar source from Gaia with unique 2MASS match \\  
    ...       &  ...   & tmmgaia & stellar source from Gaia without unique 2MASS match \\   
    ...       &  ...   & hotsubdwarf  & stellar source is the hot subdwarf list \\
    ...       &  ...   & gicycle1  & stellar source is the GI cycle 1 program \\
    ...       &  ...   & astroseis & stellar source from the \tmag\ asteroseismology task group \\
    
16            & Posflag     & ... &  Flag to identify the source of the object's position\\
    ...       &  ...   & gaia2  & stellar source from Gaia DR2 \\
    ...       &  ...   & hip  & stellar source is hipparcos \\
    ...       &  ...   & cooldwarfs  & stellar source is the cool dwarf list \\
    ...       &  ...   & 2mass  & stellar source is 2MASS \\
    ...       &  ...   & lepine  & stellar source is Lepines All-sky Catalog of Bright M Dwarfs (2011) \\
    ...       &  ...   & tmgaia & stellar source from Gaia with unique 2MASS match \\  
    ...       &  ...   & tmmgaia & stellar source from Gaia without unique 2MASS match \\   
    ...       &  ...   & hotsubdwarf  & stellar source is the hot subdwarf list \\
    ...       &  ...   & gicycle1  & stellar source is the GI cycle 1 program \\
    ...       &  ...   & 2MASSEXT  & extended source from 2MASS extended source catalog \\
    ...       &  ...   & astroseis & stellar source from the \tmag\ asteroseismology task group \\
    
21            & PMFlag & ... & Flag to identify the source of the object's proper motion\\
    ...       &  ...   & gaia2  & proper motions from \textit{Gaia} DR-2\\
    ...       &  ...   & ucac4  & proper motions from UCAC4 \\
    ...       &  ...   & tgas  & proper motions from Tycho2-\textit{Gaia} Astrometric Solution \\
    ...       &  ...   & sblink  & proper motions from SuperBlink \\
    ...       &  ...   & tycho2  & proper motions from Tycho2 \\
    ...       &  ...   & hip  & proper motions from Hipparcos \\
    ...       &  ...   & ucac5  & proper motions from UCAC5 \\
    ...       &  ...   & hsoy & proper motions from Hot Stuff for One Year \\
        
24            & PARFlag & ... & Flag to identify the source of the object's parallax\\
    ...       &  ...    & gaia2 & parallax from \textit{Gaia} DR-2\\
    ...       &  ...    & tgas  & parallax from Tycho2-\textit{Gaia} Astrometric Solution  \\
    ...       &  ...    & hip   & parallax from Hipparcos \\    

63            & TESSFlag    & ... & Flag to identify the source of the object's \tmag\ magnitude\\
    ...       &  ...   & goffs & magnitude calculated from offset with \textit{Gaia} magnitude\\
    ...       &  ...    & gpbr & \textbf{magnitude calculated from}  \\
    ...       &  ...    & gbpbrp & magnitude calculated  from  \textit{Gaia} $G_{BP}-G_{RP}$ color\\
    ...       &  ...    & rered &\textbf{magnitude calculated from} \\
    ...       &  ...    & hotsd & magnitude adopted from hot subdwarf list\\
    ...       &  ...    & cdwrf & magnitude from cool dwarf list \citep{Muirhead:2018}\\
    ...       &  ...    & gaiak & magnitude calculated from $G$ and 2MASS $K_S$\\
    ...       &  ...    & gaiaj & magnitude calculated from $G$ and 2MASS $J$ \\
    ...       &  ...    & joffset2 & magnitude calculated from 2MASS $J$ and an offset (+1.75 for $J-K_S > 1$)\\
    ...       &  ...    & hipvmag & magnitude calculated Hipparcos $V$ magnitude\\
    ...       &  ...    & gaiaoffset & magnitude calculated from $G$ and an offset\\
    ...       &  ...    & hoffset & magnitude calculated from 2MASS $H$ offset\\
    ...       &  ...    & vjh & magnitude calculated from $V$ and 2MASS $J-H$\\
    ...       &  ...    & jhk & magnitude calculated from 2MASS $J-K_S$\\
    ...       &  ...    & vjk & magnitude calculated from $V$ and 2MASS $J-K_S$\\
    ...       &  ...    & hotsubdwarf & magnitude adopted from hot subdwarf list\\
    ...       &  ...    & vk & magnitude calculated from $V$ and 2MASS $K_S$\\
    ...       &  ...    & joffset & magnitude calculated from 2MASS $J$ offset (+0.5 for $J-K_S < -0.1$)\\
    ...       &  ...    & gaiav & magnitude calculated from $G$ and $V$\\
    ...       &  ...    & tmvk & magnitude calculated from $V$ and 2MASS $K_S$ (same as vk)\\
    ...       &  ...    & from$\_$apass$\_$i & magnitude from cool dwarf list \citep{Muirhead:2018}\\
    ...       &  ...    & from$\_$sdss$\_$ik & magnitude from cool dwarf list \citep{Muirhead:2018}\\
    ...       &  ...    & gaiah & magnitude calculated from Gaia and 2MASS $H$\\
    ...       &  ...    & jh & magnitude calculated from 2MASS $J-H$\\
    ...       &  ...    & cdwarf & magnitude from cool dwarf list \citep{Muirhead:2018}\\
    ...       &  ...    & bpjk & magnitude calculated from photographic $ B$ and 2MASS J-$K_S$\\
    ...       &  ...    & voffset & magnitude calculated from $V$ and offset\\
    ...       &  ...    & koffset & magnitude calculated from 2MASS $K_S$ and offset\\
    ...       &  ...    & wmean$\_$vk$\_$jhk & magnitude from cool dwarf list \citep{Muirhead:2018}\\
    ...       &  ...    & lepine & magnitude from Lepine catalog\\
    ...       &  ...    & gicycle1 & magnitude from GI Cycle 1 proposal\\
    ...       &  ...    & from$\_$sdss$\_$i & magnitude from cool dwarf list \citep{Muirhead:2018}\\

64            & SPFlag      &  ... & Flag to identify the source of the object's stellar characteristics\\
    ...       &  ...   &  cdwrf & mass and radius from cool-dwarf list (see Sec.~\ref{sec:lists} \& \citet{Muirhead:2018} \\
    ...       &  ...        & hotsd  & mass and radius from the hot subdwarf list (see Sec.~\ref{sec:lists}) \\
    ...       &  ...        & gaia2  & characteristics computed from measured TGAS parallax \\
    ...       &  ...        & spec7 &  characteristics computed using the spectroscopic Torres relations \\
    ...       &  ...        & tic7 &  characteristics imported from TICv7 \\
\hline 
\caption{Brief description of flags in TICv8 and earlier TIC versions.\label{tab:flags}}
\end{longtable}
\end{center}

\subsection{Provenance Flags new to TICv8}
\begin{center}
\begin{longtable}[c]{|rlll|}
\hline 
Column & Name & Flags & Description  \\
\hline 
91            & EBVFlag & ... &  Flag to identify the source of the object's reddening \\
    ...       &  ...   & 0  & The star is closer than 100pc, no extinction applied \\
    ...       &  ...   & 1  & The reddening from \citet{Schlegel:1998} is applied. \\
    ...       &  ...   & 2  & The reddening from \citet{Green:2018} is applied. \\  
    
107            & TeffFlag & ... &  Flag to identify the source of the object's effective temperature \\
    ...       &  ...   & cdwrf  & \teff from the cool dwarf list \\
    ...       &  ...   & hotsd  & \teff from the hot subdwarf list \\
    ...       &  ...   & spect  & \teff from spectroscopic catalogs \\
    ...       &  ...   & gaia2  & \teff from \textit{Gaia} $G_{BP}-G_{RP}$ color \\
    ...       &  ...   & spec  & \teff imported from TICv7 \\
    
112            & gaiaqflag & ... &  Flag to identify the quality of the \textit{Gaia} astrometric and photometric information \\
    ...       &  ...   & -1  & insufficinet information \\
    ...       &  ...   & 0  & poor quality \textit{Gaia} information \\
    ...       &  ...   & 1  & good quality \textit{Gaia} information \\

114            & Vmagflag & ... &  Flag to identify the source of the object's V magnitude \\
    ...       &  ...   & gaia2  & V magnitude calculated from \textit{Gaia} $G_{BP}-G_{RP}$ color \\
    ...       &  ...   & ucac4  & V magnitude calculated from ucac4 magnitude (see \citet{Stassun:2018}) \\
    ...       &  ...   & tycho2v3  & V magnitude calculated from Tycho-V$_{T}$ magnitude (see \citet{Stassun:2018}) \\
    ...       &  ...   & tycho2v  & V magnitude calculated from Tycho-V$_{T}$ magnitude (see \citet{Stassun:2018}) \\
    ...       &  ...   & tycho  & V magnitude calculated from Tycho-V$_{T}$ magnitude (see \citet{Stassun:2018}) \\
    ...       &  ...   & apassdr9  & V magnitude imported from APASS DR-9 (see \citet{Stassun:2018}) \\ 
    ...       &  ...   & apass  & V magnitude imported from APASS DR-7 (see \citet{Stassun:2018}) \\
    ...       &  ...   & sblink  & V magnitude imported from SuperBlink (see \citet{Stassun:2018}) \\
    ...       &  ...   & mermil  & V magnitude imported from the Mermilloid catalog (see \citet{Stassun:2018}) \\
    ...       &  ...   & cdwarf  & V magnitude imported from the cool dwarf list (TICv6) (see \citet{Stassun:2018}) \\
    ...       &  ...   & cdwrf  & V magnitude imported from the cool dwarf list (TICv7) (see \citet{Stassun:2018}) \\
    ...       &  ...   & sirful  & V magnitude imported from the Sirful catalog (see \citet{Stassun:2018}) \\
    ...       &  ...   & hipvmag  & V magnitude calculated using Hipparcos V (see \citet{Stassun:2018}) \\
    ...       &  ...   & gaiak  & V magnitude calculated from \textit{Gaia} DR-1 $G$ and 2MASS $K_S$ (see \citet{Stassun:2018}) \\
115            & Bmagflag & ... &  Flag to identify the source of the object's B magnitude \\
    ...       &  ...   & tycho2b3  & B magnitude calculated from Tycho-B$_{T}$ magnitude (see \citet{Stassun:2018}) \\
    ...       &  ...   & tycho2b  & B magnitude calculated from Tycho-B$_{T}$ magnitude (see \citet{Stassun:2018}) \\
    ...       &  ...   & tycho  & B magnitude calculated from Tycho-B$_{T}$ magnitude (see \citet{Stassun:2018}) \\
    ...       &  ...   & apassdr9  & B magnitude imported from APASS DR-9 (see \citet{Stassun:2018}) \\ 
    ...       &  ...   & bpbj  & B magnitude calculated from 2MASS photometric B (see \citet{Stassun:2018}) \\
    ...       &  ...   & mermil  & B magnitude imported from the Mermilloid catalog (see \citet{Stassun:2018}) \\
116            & splists & ... &  Flag to identify if the object is in a specially curated list \\
    ...       &  ...   & cooldwarfs$\_$v8  & star is identified in the cool dwarfs specially curated list \\
    ...       &  ...   & hotsubdwarfs$\_$v8  & star is identified in the hot subdwarfs specially curated list \\
\hline 
\caption{Brief description of new provenance flags in TICv8. \label{tab:flagsv8}}
\end{longtable}
\end{center}

\section{CTL Filtergraph Portal\label{sec:appendix_filtergraph}}

Table~\ref{tbl:filtergraph} summarizes the contents of the enhanced CTL provided via the Filtergraph data visualization portal service at the URL \url{filtergraph.vanderbilt.edu/tess_ctl}.

\begin{footnotesize} 
\begin{center}
\begin{longtable}[c]{|c|l|}
 \multicolumn{2}{c}{Descriptions of CTL Contents}\\
 \hline
 Column name & Brief description \\
 \hline
 Right\_Ascension    &  Right Ascension of the star, equinox J2000.0, epoch 2000.0 (degrees)\\
 Declination   &  Declination of the star, equinox J2000.0, epoch 2000.0 (degrees) \\
 Tess\_mag & Calculated TESS magnitude \\
 Teff & Adopted effective temperature (K)\\
 Priority & Priority based on \tmag, radius, and flux contamination with boosts and de-boosts\\
 Radius & Stellar radius derived from photometry (\rsun) \\
 Mass & Stellar mass derived from photometry (\msun) \\
 ContamRatio & Ratio of contaminating flux to flux from the star\\
 Observed & 0 or 1 if the star has been observed already in 2-minute cadence\\
 Sector & Sector (or combination of sectors) in which the star was observed\\
 Galactic\_Long & Longitude in the Galactic coordinate frame (degrees)\\
 Galactic\_Lat & Latitude in the Galactic coordinate frame (degrees)\\
 Ecliptic\_Long & Longitude in the Ecliptic coordinate frame (degrees)\\
 Ecliptic\_Lat & Latitude in the Ecliptic coordinate frame (degrees)\\
 Parallax & The parallax of the star provided by either TGAS/Gaia or Hipparcos (mas)\\
 Distance & The distance of the star provided (pc)\\
 Total\_Proper\_Motion & Total proper motion of the star (mas/yr)\\
 V\_mag & Adopted $V$ magnitude\\
 J\_mag & 2MASS $J$ magnitude\\
 H\_mag & 2MASS $H$ magnitude\\
 $K_S$\_mag & 2MASS $K_S$ magnitude\\
 G\_mag & \textit{Gaia} magnitude\\
 u\_mag & SDSS u magnitude\\
 g\_mag & SDSS g magnitude\\
 r\_mag & SDSS r magnitude\\
 i\_mag & SDSS i magnitude\\
 z\_mag & SDSS z magnitude\\
 W1\_mag & ALLWISE W1 magnitude\\
 W2\_mag & ALLWISE W2 magnitude\\
 W3\_mag & ALLWISE W3 magnitude\\
 W4\_mag & ALLWISE W4 magnitude\\
 G\_BP & \textit{Gaia} DR-2 BP magnitude\\
 G\_RP & \textit{Gaia} DR-2 RP magnitude\\
 Hipparcos\_Number & {\it Hipparcos\/} ID\\
 Tycho2\_ID & {\it Tycho-2\/} ID\\
 2MASS\_ID & 2MASS ID\\
 TICID & ID for the star in the TESS Input Catalog\\
 Special\_Lists & Identifies whether a star in a special list\\
 Priority\_TIC4 & Priority based on the TIC-4 schema. \\
 Priority\_TIC5 & Priority based on the TIC-5 schema. \\
 Priority\_TIC6 & Priority based on the TIC-6 schema. \\
 Priority\_Non\_Contam & Priority without neighbor contamination. \\
 Priority\_No\_Boost &  Priority without sector boosting. \\
 Teff\_Src & Source of the effective temperature (see {\tt Teff} column)\\
 Teff\_Err & Error in the effective temperature (K)\\
 Teff\_Err\_Pos & Estimated positive error in the effective temperature from MC (K)\\
 Teff\_Err\_Neg & Estimated negative error in the effective temperature from MC (K)\\
 EBMV & Applied extinction\\
 EBMV\_Err & Error in extinction\\
 EBMV\_Src & Identifies source of adopted extinction\\
 StarChar\_Src & Identifies source of adopted setllar parameters\\
 Radius\_Err & Uncertainty in the radius (solar). \\
 Radius\_Err\_Pos & Estimated positive error in the stellar radius from MC (solar)\\
 Radius\_Err\_Neg & Estimated negative error in the stellar radius from MC (solar)\\
 Mass\_Err & Uncertainty in the mass (solar). \\
 Logg & Surface gravity (cgs). \\
 Logg\_Err & Uncertainty in the surface gravity (cgs). \\
 Rho & Density (solar). \\
 Rho\_Err & Uncertainty in the density (solar). \\
 Lum & Luminosity (solar). \\
 Metallicity & Stellar metallicity from spectra, if available (dex).\\
 Metallicity\_Err & Stellar metallicity error from spectra, if available (dex).\\
 Noise\_Star & Uncertainty from the star counts. \\
 Noise\_Sky & Uncertainty from the sky counts. \\
 Noise\_Contaminates & Uncertainty from the neighboring star counts. \\
 Noise\_Readout & Uncertainty in the detector readout. \\
 Noise\_Systematics & Uncertainty floor. \\
 Distance\_Err & Uncertainty in the disance (pc). \\
 Distance\_Err\_Pos & Estimated positive error in the distance from MC (pc)\\
 Distance\_Err\_Neg & Estimated negative error in the distance from MC (pc)\\
 \hline
 \caption{A basic description of all quantities found on the Filtergraph portal. \label{tbl:filtergraph}}
\end{longtable}
\end{center}
\end{footnotesize}

\end{document}